\documentclass{aa}
\usepackage{natbib}
\bibpunct{(}{)}{;}{a}{}{,} 
\usepackage{graphicx}
\usepackage{txfonts}
\newcommand{\Mdot}{\ensuremath{\dot{M}}}
\newcommand{\vinf}{\ensuremath{\varv_{\infty}}}
\newcommand{\vesc}{\ensuremath{\varv_{\rm esc}}}
\newcommand{\Msol}{\text{M}\ensuremath{_{\odot}}}
\newcommand{\Rsol}{\text{R}\ensuremath{_{\odot}}}
\newcommand{\Lsol}{\text{L}\ensuremath{_{\odot}}}

\newcommand{\FW}{\textsc{\large fastwind}}
\newcommand{\MESA}{\texttt{MESA}}
\newcommand{\Teff}{\ensuremath{T_{\rm eff}}}
\newcommand{\logg}{\ensuremath{\log{g}}}
\newcommand{\grad}{\ensuremath{g_{\text{rad}}}}
\newcommand{\dvdr}{\ensuremath{\frac{d\varv}{dr}}}
\newcommand{\tauF}{\ensuremath{\tilde{\tau}_F}}
\newcommand{\ferr}{\ensuremath{f_{\rm err}}}
\newcommand{\ferrmax}{\ensuremath{f_{\rm err}^{\text{max}}}}
\newcommand{\alphaeff}{\ensuremath{\alpha_{\rm eff}}}
\newcommand{\Dmom}{\ensuremath{\Mdot\vinf\sqrt{R_{\ast}}}}
\newcommand{\vturb}{\ensuremath{\varv_{\text{turb}}}}

\begin{document} 

   \title{New predictions for radiation-driven, steady-state mass-loss and wind-momentum from hot, massive stars}
   \titlerunning{Mass-loss and wind-momentum from hot, massive stars}

   \subtitle{II. A grid of O-type stars in the Galaxy and the Magellanic Clouds}

   \author{R. Bj\"orklund 
          \inst{1}
          \and
          J.O. Sundqvist\inst{1}\and
          J. Puls\inst{2} \and
          F. Najarro\inst{3}}
          
   \authorrunning{R. Bj\"orklund et al.}

   \institute{KU Leuven, Instituut voor Sterrenkunde, Celestijnenlaan 200D, 3001 Leuven, Belgium\\
              \email{robin.bjorklund@kuleuven.be}
              \and LMU M\"unchen, Universit\"atssternwarte, Scheinerstr. 1,
     81679 M\"unchen, Germany\and Centro de Astrobiologia, Instituto Nacional de Tecnica Aerospacial, 28850 Torrejon de Ardoz, Madrid, Spain}

   \date{Received 11 May 2020; accepted 12 August 2020}

 
  \abstract
   {Reliable predictions of mass-loss rates are important for massive-star evolution computations.}
   {We aim to provide predictions for mass-loss rates and wind-momentum rates of O-type stars, carefully studying the behaviour of these winds as functions of stellar parameters like luminosity and metallicity.}
   {We use newly developed steady-state models of radiation-driven winds to compute the global properties of a grid of O-stars. The self-consistent models are  calculated by means of an iterative solution to the equation of motion using full NLTE radiative transfer in the co-moving frame to compute the radiative acceleration. In order to study winds in different galactic environments, the grid covers main-sequence stars, giants and supergiants in the Galaxy and both Magellanic Clouds.}
   {We find a strong dependence of mass-loss on both luminosity and metallicity. Mean values across the grid are $\Mdot\sim L_{\ast}^{2.2}$ and $\Mdot\sim Z_{\ast}^{0.95}$, however we also find a somewhat stronger dependence on metallicity for lower luminosities. Similarly, the mass loss-luminosity relation is somewhat steeper for the SMC than for the Galaxy. In addition, the computed rates are systematically lower (by a factor 2 and more) than those commonly used in stellar-evolution calculations. Overall, our results agree well with observations in the Galaxy that account properly for wind-clumping, with empirical $\Mdot$ vs. $Z_\ast$ scaling relations, and with observations of O-dwarfs in the SMC.} 
   {Our results provide simple fit relations for mass-loss rates and wind momenta of massive O-stars stars as functions of luminosity and metallicity, valid in the range $T_{\rm eff} = 28000 - 45000$\,K. Due to the systematically lower $\Mdot$, our new models suggest that new rates might be needed in evolution simulations of massive stars.}

   
   \maketitle
%

\section{Introduction}

    Hot, massive stars, with masses $\gtrapprox 9 \Msol$, of spectral type O and B lose a significant amount of mass due to their radiation-driven stellar winds \citep{Castor75}. This mass loss has a dominant influence on the life-cycles of massive stars, as well as in determining the properties of the remnants left behind when these stars die \citep[e.g.][]{Smith14}. The rates at which these stars lose mass, on the order of $\Mdot \sim 10^{-5\ldots-9}$\Msol/yr, comprise a key uncertainty in current models of stellar evolution (even on the main sequence where the stars typically are 'well behaved', e.g. \citealt{Keszthelyi17}), simply because the mass of a star is the most important parameter determining its evolution. In addition to the loss of mass, also angular momentum is lost through stellar winds affecting the surface rotation speeds of these stars. Moreover, uncertainties related to mass loss have consequences on galactic scales beyond stellar physics, as massive stars provide strong mechanical and radiative feedback to their host environment \citep{Bresolin08}.
    
    It is therefore important to have reliable quantitative predictions of mass-loss rates and wind-momenta of massive stars. In the first paper of this series \citep[][from here on Paper I]{Sundqvist19}, we developed a new method to provide mass-loss predictions based on steady-state wind models using radiative transfer in the co-moving frame (CMF) to compute the radiative acceleration \grad.
    Building on that, this paper now presents results from a full grid of models computed for O-stars in the Galaxy, the Small and the Large Magellanic Cloud, analyzing the general dependence on important stellar quantities like luminosity and metallicity. Paper I showed that simulations using CMF radiative transfer suggest reduced mass-loss rates as compared to the predictions normally included in models of massive star evolution \citep{Vink00,Vink01}. Although this paper focuses on the presentation of our new rates for O-stars in different galactic environments, a key aim for future publications within this series will be to directly implement the results of the new wind models into calculations of massive-star evolution.
    
    Since most of the important spectral lines driving hot-star winds are metallic, a strong dependence on metallicity $Z_{\ast}$ is expected for the mass-loss rate \citep{Kudritzki87,Vink01,Mokiem07b}. 
    To investigate this, we here compute models tailored to our local Galactic environment, assuming a metal content like that in the Sun, as well as for the Magellanic Clouds. The metallicities of these external galaxies are about half the Galactic one for the Large Magellanic Cloud (LMC) and a fifth for the Small Magellanic Cloud (SMC). Using these three regimes we aim to study the mass-loss rate in function of $Z_{\ast}$. The Clouds are interesting labs for stellar astrophysics because the distances to the stars are relatively well constrained, providing values of their luminosities and radii.
    Another reason we are focusing on the Clouds is that quantitative spectroscopy of individual stars there has been performed and compiled into an observed set of scaling-relations for wind-momenta \citep{Mokiem07b}. While quantitative spectroscopy of individual hot, massive stars nowadays is possible also in galaxies further away \citep[e.g.][]{Garcia19}, such studies are only in their infancy.
    
    In order to derive global dependencies and relations for the mass-loss rates and wind-momenta, we perform a study  using a grid of O-star models. Thanks to the fast performance of the method, as explained in detail in Paper I, this is finally possible for hydrodynamically consistent steady-state models with a non-paramterized CMF line-force computed without any assumptions about underlying line-distribution functions. In Section \ref{sec:Methods} we briefly review our method for computing mass-loss rates, highlighting one representative model from the grid. In Section \ref{sec:Results} the results of the full grid of models are shown, first for the Galaxy and then including the Magellanic Clouds, in terms of computed wind-momenta and mass-loss rates. From these results we derive simple fit relations for the dependence on luminosity and metallicity. Section \ref{sec:Discussion} provides a discussion of the results, highlighting the general trends and comparing to other existing models and to observations. Additionally we address the implication for stellar evolution and current issues like the so-called "weak wind" problem. Section \ref{sec:Conclusions} contains the conclusions and future prospects.

\section{Methods}\label{sec:Methods}

    A crucial part about the radiation-driven steady-state wind models used in our research is that they are hydrodynamically consistent. This means that the equation of motion (e.o.m.) in the spherically symmetric, steady-state case is solved as described in Paper I. This e.o.m. reads
    \begin{align}\label{eom}
        \varv(r)\frac{d\varv(r)}{dr}\left(1-\frac{a^2(r)}{\varv^2(r)}\right) = \grad(r)-g(r)+\frac{2a^2(r)}{r}-\frac{da^2(r)}{dr}.
    \end{align}
    Here $\varv(r)$ is the velocity, $a(r)$ the isothermal sound speed, $\grad(r)$ the radiative acceleration, $g(r)=GM_{\ast}/r^2$ the gravity, with gravitation constant $G$, $M_{\ast}$ the stellar mass used in the model and $r$ the radius coordinate. The temperature structure $T(r)$ enters the equation through the isothermal sound speed
    \begin{align}
        a^2(r)=\frac{k_bT(r)}{\mu(r) m_H},
    \end{align}
    with $k_B$ Boltzmann's constant, $\mu(r)$ the mean molecular weight and $m_H$ the mass of a hydrogen atom. Equation \ref{eom} has a sonic point where $\varv(r) = a(r)$. Because in the above formulation \grad\ is only an explicit function of radius, and not of the velocity gradient, the corresponding critical point in the e.o.m. is this sonic point (see also discussion in Paper I). The radiative acceleration also depends on velocity and mass loss, of course, but in our method these dependencies are  implicit;
    they are accounted for through the iterative updates of the velocity and density structure, and do not affect the critical point condition in an explicit way.
    
    For given stellar parameters luminosity $L_{\ast}$, mass $M_{\ast}$, radius $R_{\ast}$ (to be defined below), and metallicity $Z_{\ast}$, the e.o.m. \eqref{eom} is solved to obtain $\varv(r)$ for the subsonic photosphere and supersonic radiation-driven wind. For a steady-state mass-loss rate \Mdot\ the mass conservation equation $\Mdot=4\pi r^2\rho\varv$ gives the density structure $\rho(r)$. The wind models further rely on the NLTE (non-local thermodynamic equilibrium) radiative transfer in \FW\ (see Paper I and \citealt{Puls17}) for the computation of \grad, by means of a co-moving frame (CMF) solution and without using any parametrised distribution functions.
    The atomic data are taken from the \textsc{\large{WM-basic}} data base \citep{Pauldrach01}. This compilation consists of more than a million spectral lines, including all major metallic elements up to Zn and all ionisation stages relevant for O-stars. The data base is the same as the one utilized in Paper I, as well as in previous versions of the {\sc fastwind} code \citep[see, e.g.,][]{Puls05}. Also the hydrogen and helium model atoms are identical to those used in our previous work. We note that since \textsc{\large{WM-basic}} has been calibrated for diagnostic usage in the UV regime, its principal data base should be ideally suited for the radiative force calculations in focus here. In the NLTE and \grad\ calculation, we account for pure Doppler-broadening alone\footnote{Relaxing this assumption would have an only marginal effect on the occupation numbers \citep[e.g.,][]{Hamann81,Lamers87}. Also for the calculation of \grad, pure Doppler-broadening is sufficient, since (i) only few strong lines have significant natural and -- in the photosphere -- collisionally line-broadened wings that could contibute to \grad; and (ii), because of line-overlap effects, these wings are typically dominated by the (Doppler-core) line opacity from other transitions, which then dominate the acceleration.} as depth and mass dependent profiles, including also a fixed microturbulent velocity \vturb\ (see further Paper I, and Sects. 2.3 and 4.5 of this paper). 
    More specific details of the NLTE and radiative transfer in the new CMF \FW\ v11 have been laid out in detail in \citet[][see also \citet{Puls17} and Paper I]{Puls20}.
    
    The steady-state $\dot{M}$ and $\varv(r)$ are converged in the model computation starting from a first guess. For all simulations presented in this paper, the start-value for $\dot{M}$ is taken as the mass-loss rate predicted by the \citet{Vink01} recipe using the stellar input parameters of the model. The initial velocity structure is obtained by assuming that a quasi-hydrostatic atmosphere connects at $\varv_{\rm tr} \approx 0.1a(T=\Teff)$ to a so-called $\beta$-velocity law
    \begin{align}
        \varv(r) = \vinf\left(1-b\frac{R_{\ast}}{r}\right)^{\beta},
    \end{align}
    with \vinf \, the terminal wind speed, $R_{\ast}$ the stellar radius, $\beta$ a positive exponent and $b$ a constant derived from the transition velocity $\varv_{\rm tr}$. We further define the stellar radius
    \begin{align}\label{Rstar}
        R_{\ast} \equiv r(\tauF = 2/3),
    \end{align}
    where $\tauF$ is the spherically modified flux-weighted optical depth 
    \begin{align}
        \tauF(r) = \int_{r}^{\infty}\rho(\hat{r})\kappa_F(\hat{r})\left(\frac{R_{\ast}}{\hat{r}}\right)^2d\hat{r},
    \end{align}
    for the flux-weighted opacity $\kappa_{\rm F}$ $\rm [cm^2/g]$. The flux-weighted opacity is 
    related to the radiative acceleration as $g_{\rm rad} = \kappa_{\rm F} L_\ast/(4 \pi c r^2)$. 
    After each update of the hydrodynamical structure (see below), an NLTE/radiative transfer loop is carried out, to converge the occupation numbers and \grad. The velocity gradient at the wind critical point is then computed by applying l'H\^opital's rule, after which the momentum equation \eqref{eom} is solved with a Runge-Kutta method to obtain the velocity structure $\varv(r)$ above and below the critical point. This integration is performed by shooting both outwards from the sonic point to a radius of about 100$R_{\ast}$ to 140$R_{\ast}$ and inwards towards the star, stopping at $r=r_\text{min}$ when a column mass $m_c = \int_{r_\text{min}}^{\infty}\rho(r) dr = 80$ g/cm$^2$ is reached.
    
    On top of the velocity structure, also the temperature structure is updated every hydrodynamic iteration. We use a simplified method similar to \citet{Lucy71} to speed up the convergence; the temperature throughout the radial grid is calculated as
    \begin{align}\label{Lucy}
        T(r) = \Teff\left(W(r) + \frac{3}{4}\tauF(r)\right)^{1/4},
    \end{align}
    where $W(r)$ is the dilution factor given by
    \begin{align}
        W(r) = \begin{cases}
        \frac{1}{2}\left(1-\sqrt{1-\frac{R_{\ast}^2}{r^2}}\right) & \quad \text{if } r>R_{\ast}, \\
        \frac{1}{2} & \quad \text{if } r\leq R_{\ast}.
        \end{cases}
    \end{align}
    The effective temperature \Teff\ in equation \eqref{Lucy} is defined as $\sigma\Teff^4\equiv L_{\ast}/4\pi R_{\ast}^2$, with $\sigma$ the Stefan-Boltzman constant and $R_{\ast}$ as defined in equation \eqref{Rstar}. Additionally there is a floor temperature of $T\approx0.4\Teff$ like in previous versions of \FW. This temperature structure is held fixed during the following NLTE iteration, meaning that, formally,  perfect radiative equilibrium is not achieved; however, the effects from this on the wind dynamics are typically negligible (see Paper I).
    
    As described in Paper I, the singularity and regularity conditions applied in CAK-theory cannot be used to update the mass-loss rate, because in the approach considered here \grad\ does not explicitly depend on density, velocity, or the velocity gradient. Instead, here at iteration $i$ the mass-loss rate used in iteration $i+1$ is updated to counter the current mismatch in the force balance at the critical sonic point (see also Sander et al. 2017). For a hydrodynamically consistent solution the quantity
    \begin{align}
        f_{\text{rc}} = 1-\frac{2a^2}{rg}+\frac{da^2}{dr}\frac{1}{g}
    \end{align}
    should be equal to $\Gamma=\grad/g$ at the sonic point. In order to fulfill the e.o.m. \eqref{eom} the current mismatch is countered by updating the mass-loss rate according to $\Mdot_{i+1} = \Mdot_i(\Gamma/f_{\text{rc}})^{1/b}$, following the basic theory of line-driven winds where $\grad\ \propto 1/\Mdot^{b}$ \citep{Castor75}. Our models take a value of $b=1$ in the iteration loop, providing a stable way to converge the steady-state mass-loss rate. From this \Mdot\ and the computed velocity field, a new density is obtained from the mass conservation equation.
    
    \subsection{Convergence}\label{Convergence}
        
        The new wind structure ($\varv(r), \rho(r), T(r), \Mdot$) is used in the next NLTE iteration loop to converge the radiative acceleration once more in the radiative transfer scheme. This method is then iterated until the error in the momentum equation is small enough to consider the model as converged, and thus hydrodynamically consistent. By rewriting the e.o.m. \eqref{eom} the quantity describing the current error is
        \begin{align}
            \ferr(r) = 
            1-\frac{\Lambda}{\Gamma},
        \end{align}
        with
        \begin{align}
            \Lambda = \frac{1}{g}\left(\varv\dvdr\left(1-\frac{a^2}{\varv^2}\right)+g-\frac{2a^2}{r}+\frac{da^2}{dr}\right),
        \end{align}
        for each radial position $r$. For a hydrodynamically consistent model \ferr\ is zero everywhere and $\Gamma$ should thus be equal to $\Lambda$. As such, in the models we need the maximum error in the radial grid
        \begin{align}
            \ferrmax = \max{\left(\text{abs}\left(\ferr\right)\right)}
        \end{align}
        to be close enough to zero; in this paper we require a threshold 0.01, meaning the converged 
        model is dynamically consistent to within a percent.
        Additional convergence criteria apply to the mass loss rate and the velocity structure. These quantities are not allowed to vary by more than $2\%$ for the former and $3\%$ for the latter between the last two hydrodynamic iteration steps reaching convergence. 
    
    \subsection{Generic model outcome}\label{Model_outcome}

\begin{figure}
            \centering
            \includegraphics[width=\hsize]{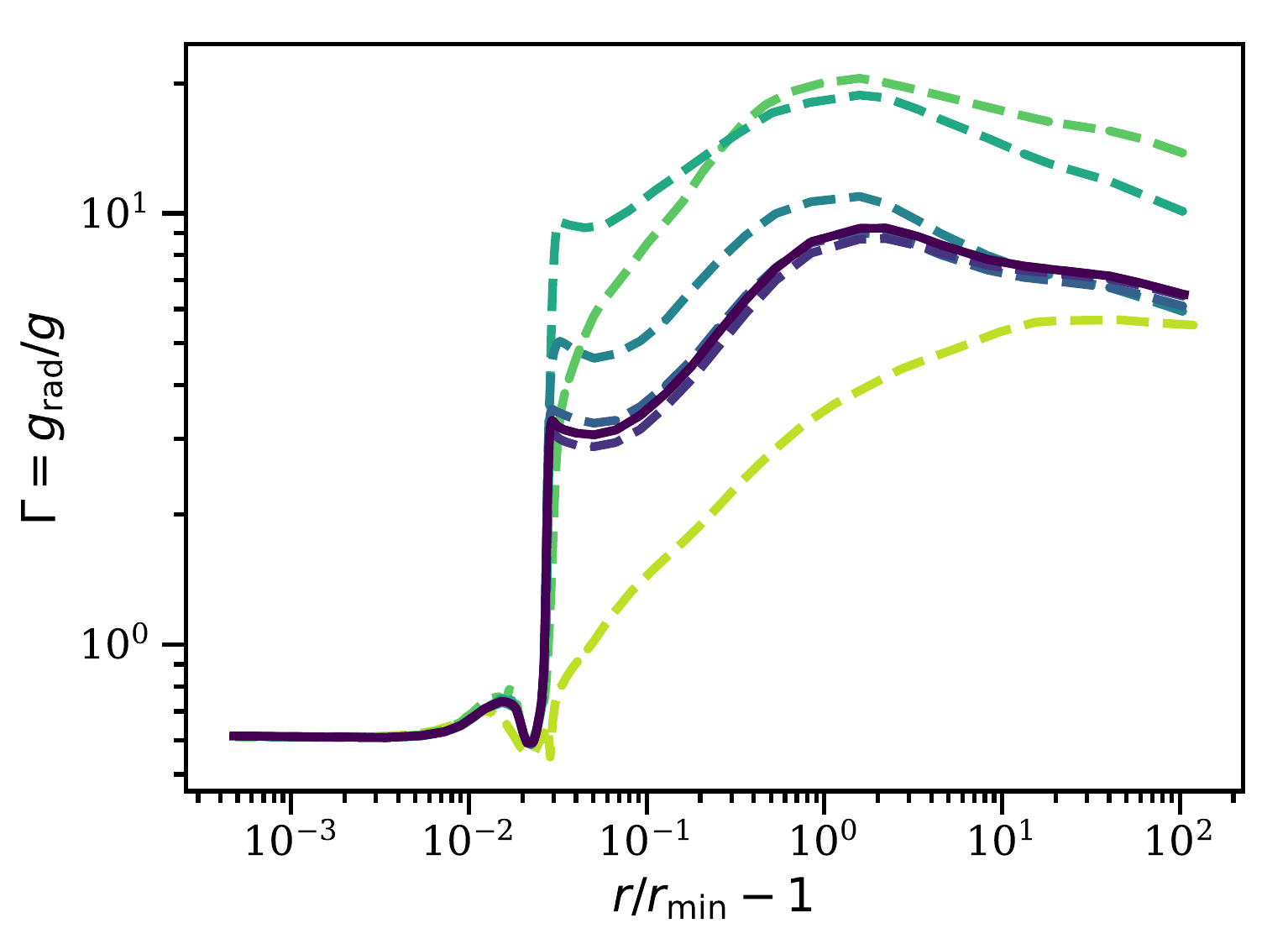}
            \includegraphics[width=\hsize]{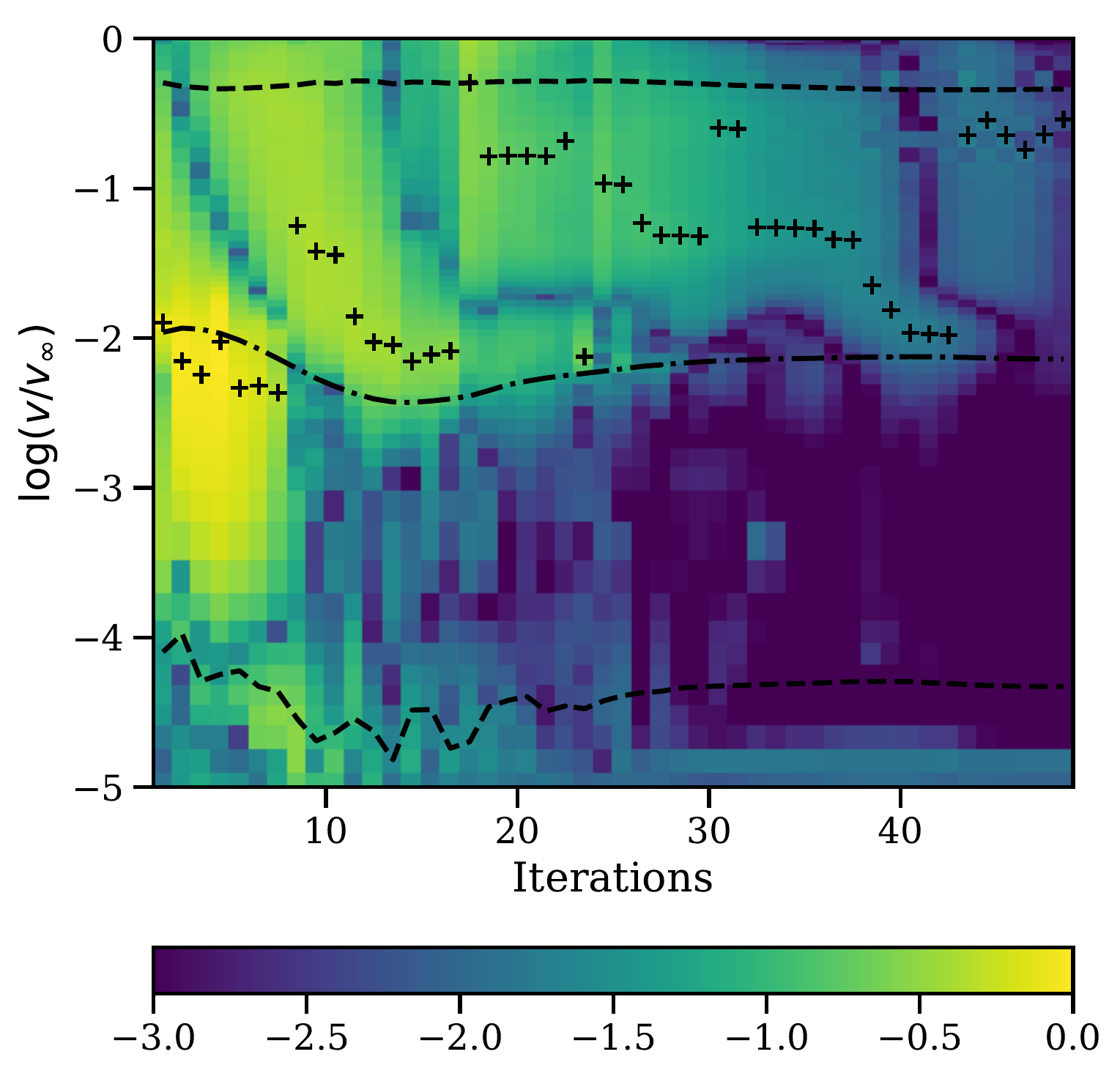}
            \caption{Top panel: Value of $\Gamma$ versus scaled radius-coordinate for 7 (non-consecutive) hydrodynamic iterations over the complete run. The starting structure (yellow) relaxes to the final converged structure (dark blue). Bottom panel: Color map of $\log(\ferr)$ for all hydrodynamic iterations; on the abscissa is hydrodynamic iteration number and on the ordinate scaled wind velocity. The pluses indicate the location of $\ferrmax$ for each iteration, the dashed lines the limits between which $\ferrmax$ is computed and the dash-dotted line the location of the sonic point.}
            \label{iters}
        \end{figure}

        As a first illustration, some generic outcomes of one characteristic simulation are now highlighted; the model parameters are listed in Table \ref{table:param}.
        \begin{table}
            \caption{Parameters of the characteristic model as described in Section \ref{Model_outcome}. The parameter \logg$_\ast$\ introduced here is the logarithm of the surface gravity which is $g(R_{\ast}) = g_\ast$.}
            \centering
            \begin{tabular}{c c c c c c c}
            \hline\hline
            $M_\ast\ $ & $R_\ast\ $ & \Teff\ & \logg$_\ast$\ & \Mdot\ & \vinf\ & $Z_{\ast}$  \\
            $ [\Msol]$ & $ [\Rsol]$ &  [K] &  & [$\Msol$/yr]& [km/s] & [$Z_{\odot}$] \\
            \hline
             58 & 13.8  & 44616  & 3.92 & $1.4\times10^{-6}$ & 3152 & 1.0  \\
            \hline
            \end{tabular}
            \label{table:param}
        \end{table}
        The top panel of Figure \ref{iters} shows the evolution of $\Gamma$ for several iterations in the scheme starting from an initial $\beta$-law structure.
        The characteristic pattern of a steep wind acceleration starting around the sonic point, where $\Gamma\approx1$, is clearly visible throughout the iteration loop. The model starts off being far from consistency (yellow), but as the error in the hydrodynamical structure becomes smaller the solution eventually relaxes to a final converged velocity structure (dark blue). The innermost points deepest in the photosphere remain quite constant, since the deep photospheric layers relax relatively quickly. In the bottom panel of Figure \ref{iters} a color plot of the iterative evolution of the model error \ferr\ throughout the wind can be seen. The point of maximum error at each iteration is marked with a plus, and the dash-dotted line shows where the velocity equals the sound speed. The dashed lines further show the boundaries within which \ferrmax\ is computed, where we note that the part at very low velocity is excluded because here the opacity is parametrised (see below). In addition a few of the outermost points are formally excluded in the calculation of \ferrmax\ (due to resolution considerations).
        Since the calculation of \ferrmax\ excludes the innermost region and a few outermost points, these points do not contribute to the condition of convergence based on the error in \ferr. Nonetheless the models do provide reliable terminal wind speeds, as they additionally require the complete velocity structure (including \vinf) to be converged to better than $3\%$ between the final two iteration steps (see above). The figure illustrates explicitly how both the overall and maximum errors generally decrease throughout the iteration cycle of the simulation. We note that, after some initial relaxation, for this particular model the position of maximum error always lies in the supersonic region, often quite close to the critical sonic point.
        
\begin{figure}
            \includegraphics[width=\hsize]{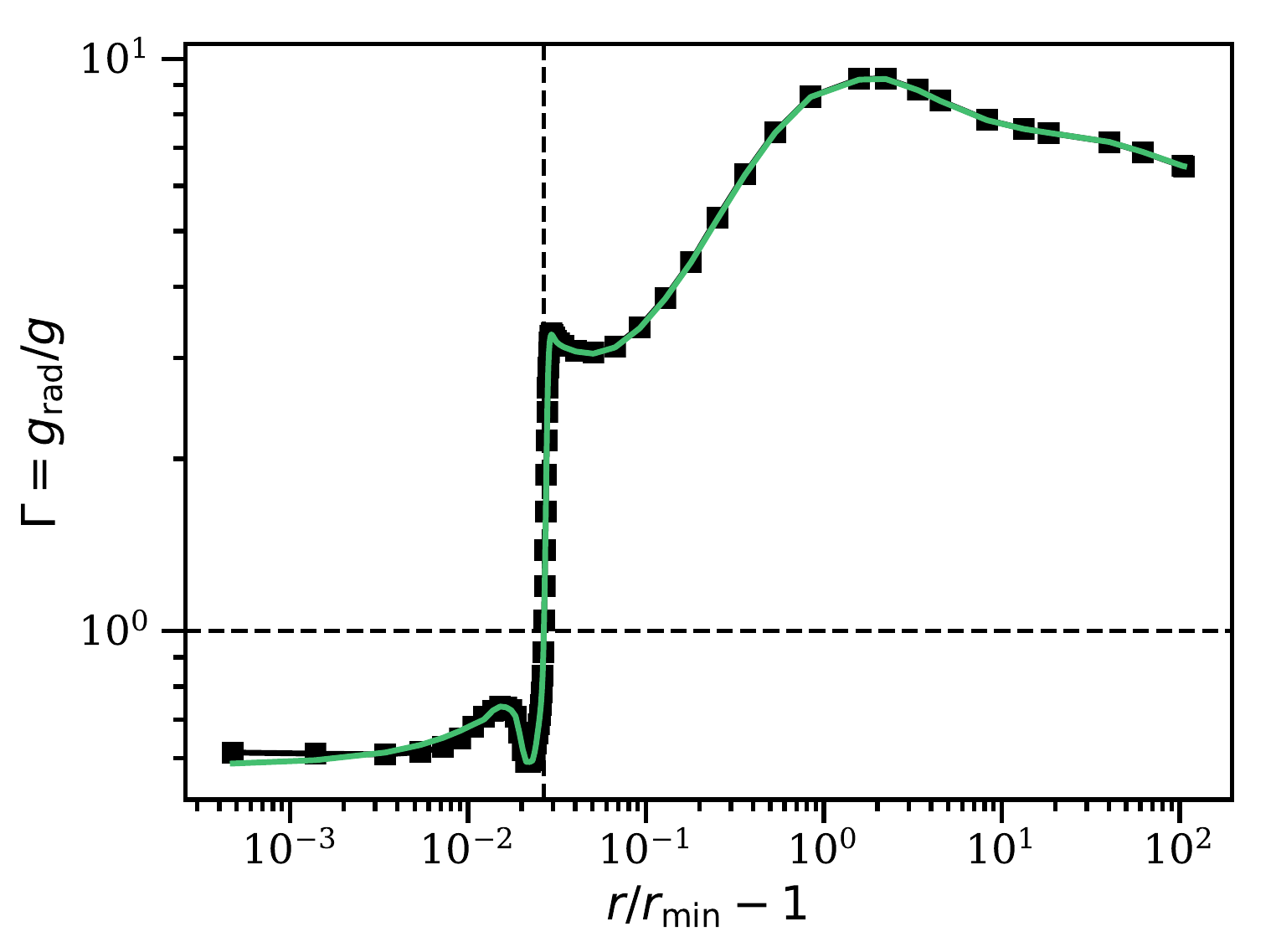}
            \includegraphics[width=\hsize]{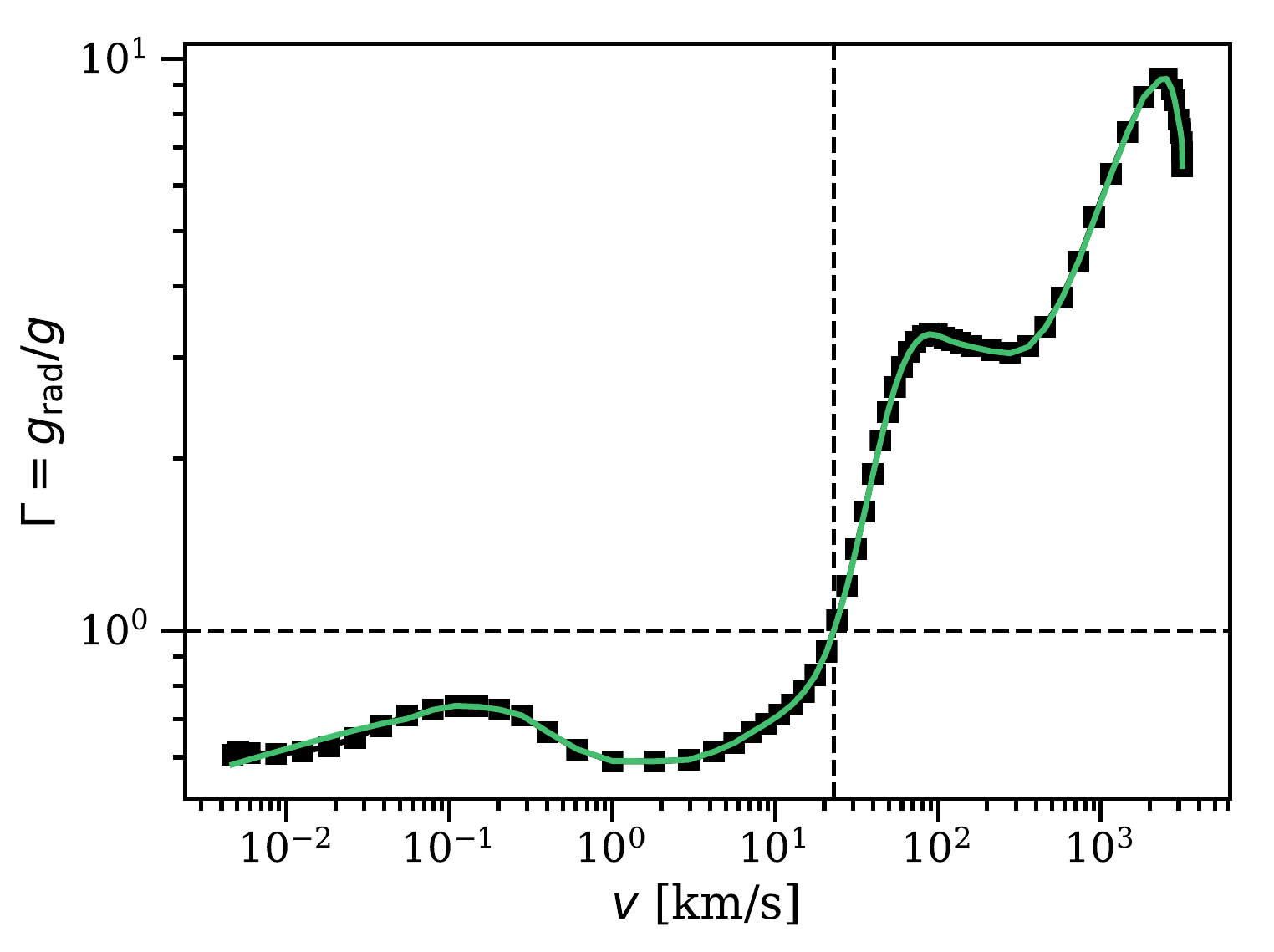}
            \caption{Top panel: Final converged structure of the characteristic model showing $\Gamma$ in black squares and $\Lambda$ (see text for definition) as a green line.  The black dashed lines show the location of the sonic point approximately at $\Gamma=1$ (but not exactly because of the additional pressure terms). Bottom panel: Same as in the top panel, however versus velocity (which resolves the inner wind more).}
            \label{Gamma_rest}
        \end{figure}

        At the end of the sequence, the model is dynamically consistent and $\Gamma$ matches the other terms in the equation of motion $\Lambda$. This is illustrated in Figure \ref{Gamma_rest}.
        This figure compares $\Gamma$ and $\Lambda$ at each radial point of the converged model, showing a clear match between the quantities. Only below a velocity $\varv\lesssim0.1$ km/s there is some discrepancy; this arises because in these quasi-static layers the flux-weighted opacity is approximated by a Kramer-like parametrisation (see Paper I), which is useful to stabilise the base in the deep subsonic atmosphere. It is important to point out that this parametrisation is applied only at low velocities and high optical depths, and so does not affect the structure of the wind or the derived global parameters.
        
\begin{figure}
            \centering
            \includegraphics[width=\hsize]{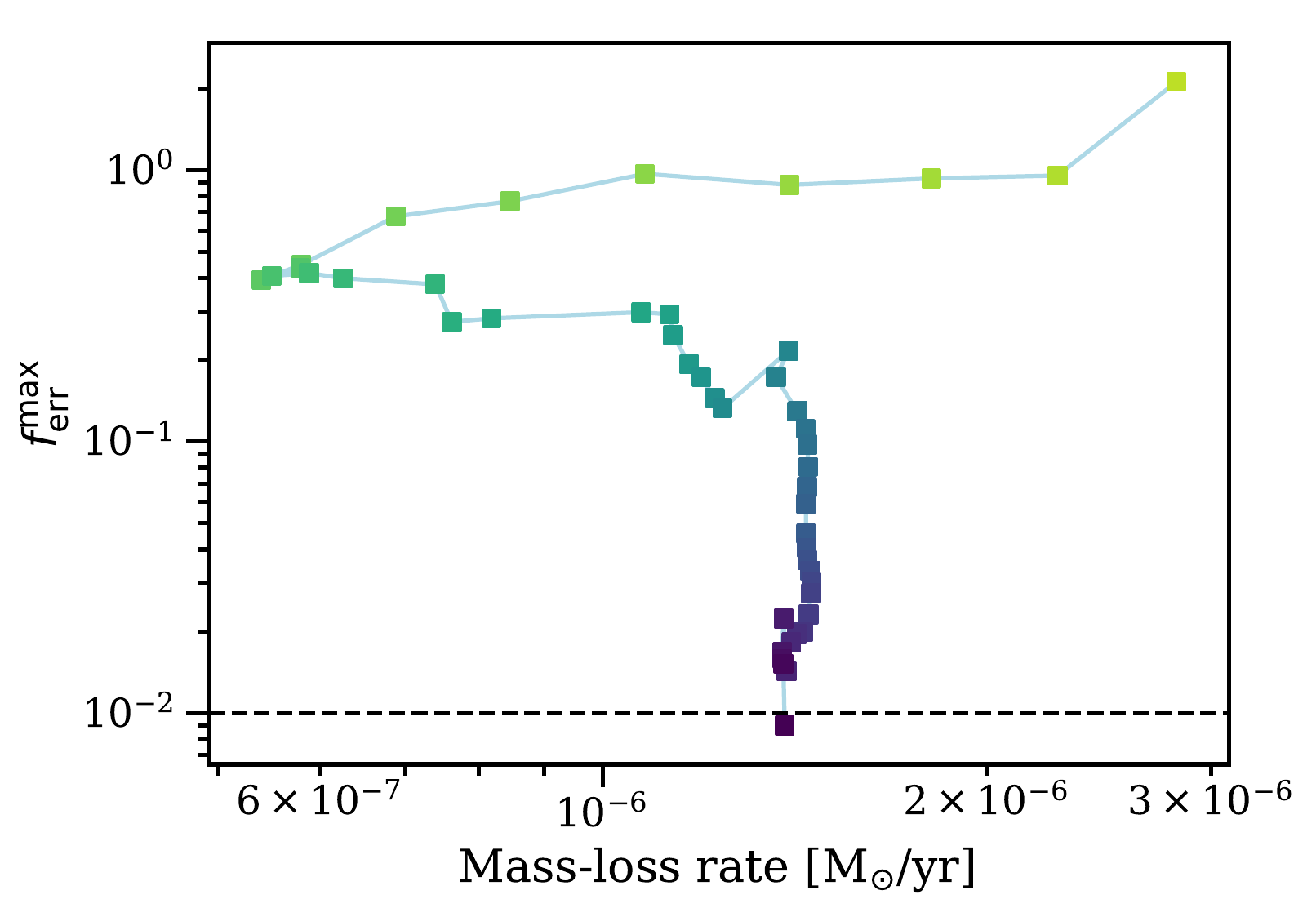}
            \includegraphics[width=\hsize]{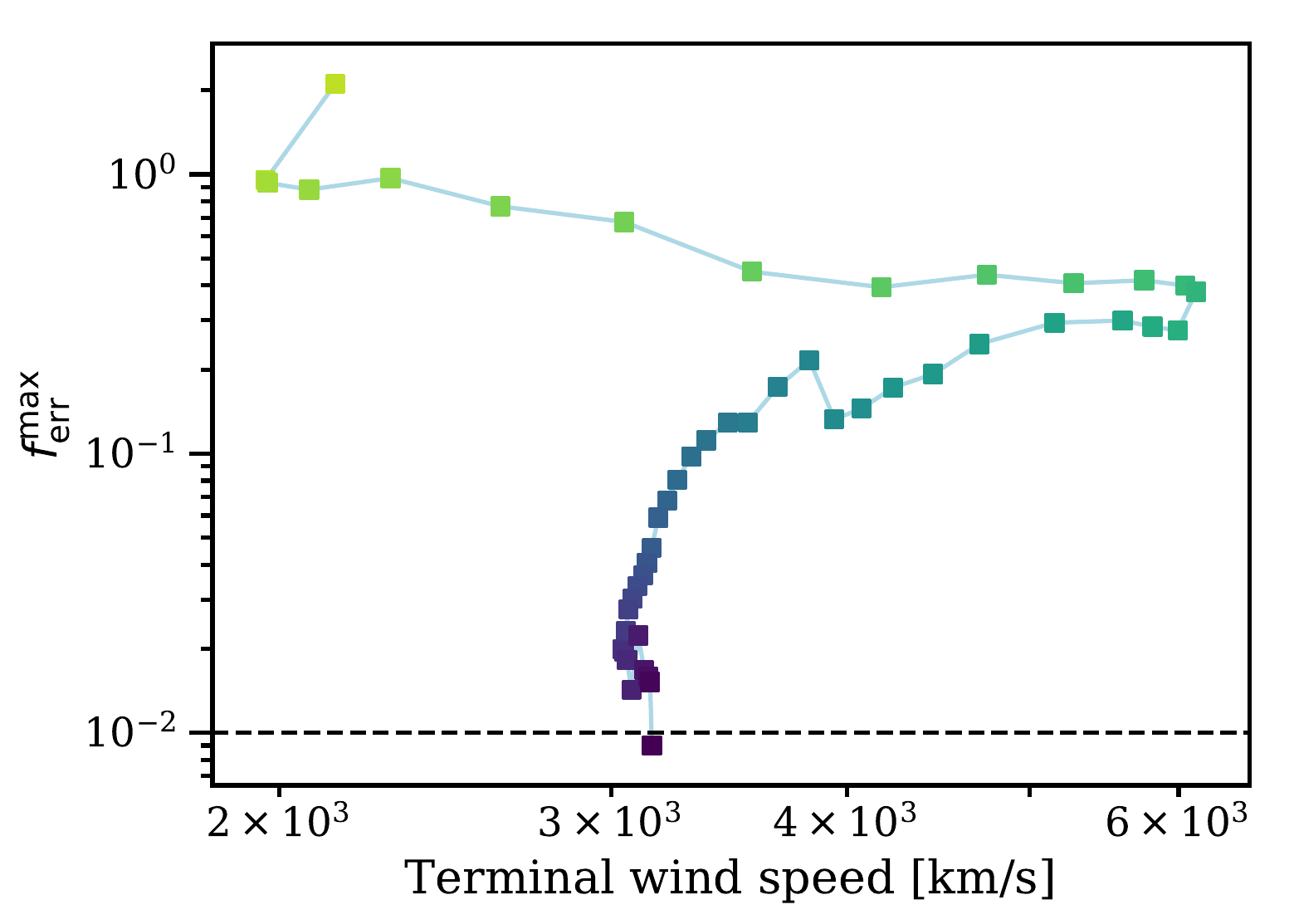}
            \caption{Top panel: Iterative behaviour of the mass-loss rate as \ferrmax\ decreases toward a value below 1\%. The color signifies the iteration number starting from $\Mdot$ as predicted by the Vink et al recipe in light green. Bottom panel: Iterative behaviour of the terminal velocity \vinf\ towards convergence.}
            \label{conv}
        \end{figure}

\begin{figure}
            \centering
            \includegraphics[width=\hsize]{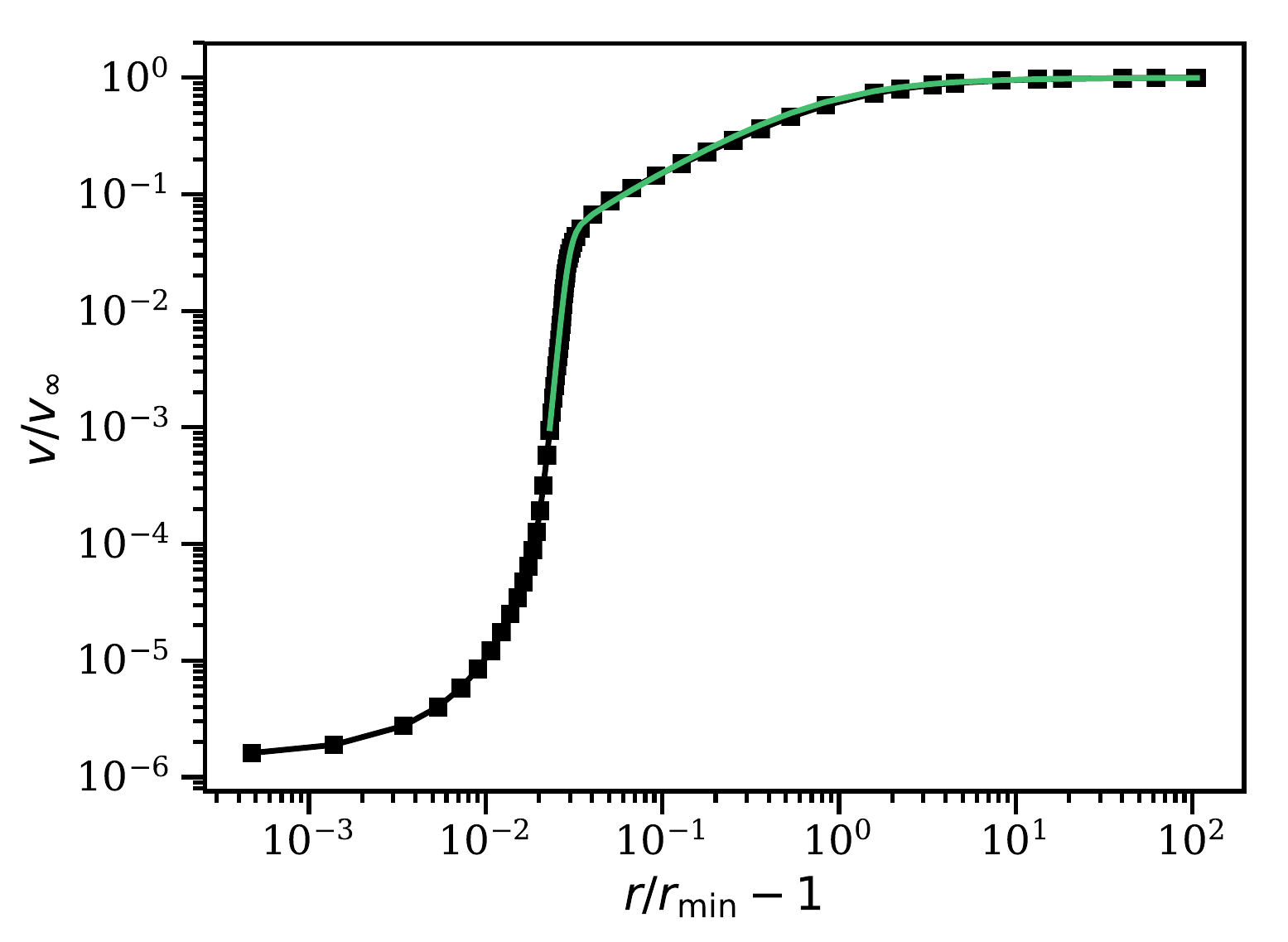}
            \caption{Converged velocity structure for the characteristic model of Section \ref{Model_outcome}, showing velocity over terminal wind speed versus the scaled radius-coordinate in black. The green line shows a fit using a double $\beta$-law following equation \eqref{2beta} (see text for details).}
            \label{vr}
        \end{figure}

        The behaviour of the mass loss rate \Mdot\ is important to understand for the purposes of this paper. The top panel of Figure \ref{conv} shows \ferrmax\ of the model for all iterations versus the mass-loss rate computed for that iteration.
        The general trend is that \ferrmax\ decreases quite consistently during the iteration cycle. The mass loss rate can be seen to converge to one value as the structure gets closer to dynamical consistence. Indeed, in the last couple of iterations \Mdot\ only changes minimally from its former value.  In the bottom panel of Figure \ref{conv} the same plot is shown for the iterative evolution of the terminal wind speed. Also this quantity 
        displays a stable convergence behaviour towards one final value. The quantities \Mdot\ and \vinf\ are from Figure \ref{conv} seen to have an anti-correlation, as for this model the total wind-momentum rate \Mdot\vinf\ does not vary much after the first few iterations.
        Finally, Figure \ref{vr} shows the converged velocity structure versus the modified radius coordinate $\frac{r}{r_{\text{min}}}-1$, where $r_{\text{min}}$ is the inner-most radial point of the grid.
        This figure illustrates the very steep acceleration around the sonic point that is 
        characteristic for our O-star models, followed by a supersonic $\beta$-law-like 
        behaviour typical of a radiation-driven stellar wind. In the figure we add a fit to the velocity structure using a double $\beta$-law defined by equation \eqref{2beta}, and elaborate further on this comparison in Section \ref{beta}.   
    
    \subsection{Grid setup}
        
        In order to study the mass-loss rates of O stars in a quantitative way, a model-grid is constructed by varying the fundamental input stellar parameters. For the Galactic stars, we use the calibrated stellar parameters obtained from a theoretical \Teff\ scale of a set of O stars from \citet{Martins05} (adopting the values from their Table 1). The Martins et al. parameters were retrieved by means of a grid of non-LTE, line-blanketed synthetic spectra models using the code CMFGEN \citep{Hillier98}. Here, a self-consistent wind model as described above is calculated for the stellar parameters of each star in the grid, resulting in predictions of the wind structure, terminal velocity and mass loss rate. The wind models are computed with the microturbulent velocity kept at a standard value for O-stars $\vturb=10$ km/s (see also Paper I), and the helium number abundance is kept fixed at $Y_{\text{He}}=n_{\text{He}}/n_{\text{H}}=0.1$. Moreover, the simulations are performed without any inclusion of clumping or high-energy X-rays. As discussed in detail in Paper I, while such wind clumping is both theoretically expected \citep{Owocki88,Sundqvist18,Driessen19} and observationally established (see Sect. 4.2), it is still uncertain what effect this might have on theoretically derived global mass-loss rates (indeed, in the simple tests performed in Paper I the effect was only marginal). In addition, any inclusion of wind clumping into steady-state models will inevitably be of ad-hoc nature (see also discussion in Paper I).
        A study by \citet{Muijres11}, for example, shows that introducing clumps can sometimes change their predicted mass loss rate 
        by as much as an order of magnitude. The models used in their study, however, use a global energy constraint to derive the mass loss rate for an assumed fixed $\beta$ velocity law. As such, they are not locally consistent and thus might not necessarily fulfill the force requirements around the sonic point. By contrast, the models presented here are (by design) both locally and globally consistent, with a mass loss rate that is primarily sensitive to the conditions around the critical sonic point. It might only be influenced if these corresponding regions where \Mdot\ is determined, are strongly clumped. This is a key difference between the mass-loss rates derived here and those in \citet{Muijres11}, and a prime reason that, contrary to their findings, \Mdot\ in our models does not seem to change significantly when introducing clumping in the supersonic parts. However, as also discussed in Paper I, the terminal velocities are typically affected by adding such clumping. Namely, since \grad\ is altered in the supersonic regions due to the presence of the clumps, this can lead to modified values of \vinf.
        So even though \Mdot\ is barely influenced when including typical wind-clumping, it remains an uncertainty of the current models, and future work should aim for a more systematic study of possible feedback effects from clumping upon also the steady-state e.o.m.
        
        In any case, the models presented here contain 12 spectroscopic dwarfs, 12 giants and 12 supergiants for each value of metallicity $Z_{\ast}$. For simplicity, the same stellar parameters ($L_\ast$, $M_\ast$, $R_\ast$) as for the Galactic grid were assumed to create models for the LMC and SMC, changing only their metallicity. This set-up has the advantage of enabling a rather direct comparison for the model-dependence on metallicity, independent of the other input parameters. The used metallicities in the grid are $Z_{\text{Galaxy}}=Z_{\odot}$, $Z_{\text{LMC}} = 0.5 Z_{\odot}$ and $Z_{\text{SMC}} = 0.2 Z_{\odot}$, respectively. The value of the Solar metallicity is here taken to be $Z_{\odot}=0.013$ \citep{Asplund09}. In total this gives 108 models, with input stellar parameters as listed in table \ref{all_param} in Appendix \ref{appendix}.


\section{Results}\label{sec:Results}

    The results for all 108 models are added to table \ref{all_param}, containing the derived values for \Mdot\ and \vinf. The runs typically took about 50 iterative updates of the hydrodynamical structure to converge, where for most parts the corresponding calculation of $g_{\rm rad}$ (aside from the first ones) takes 10 to 15 NLTE radiative transfer steps per hydrodynamic update. Using the criteria as presented in Section \ref{Convergence}, all models presented in this paper are formally converged.
    
    The following subsections highlight the results for the Galaxy, the Small and the Large Magellanic Clouds.
    

    \subsection{The Galaxy}\label{results_G}
    
        When studying the overall behaviour of line-driven winds, it is useful to look at a modified wind-momentum rate as function of the stellar luminosity:  
        \begin{equation}\label{WMLR}
            \Dmom \propto L_{\ast}^{x}
        \end{equation}
        where the left hand side is the so-called modified wind-momentum rate $D_{\text{mom}} \equiv \Dmom$ \citep{Kudritzki95,Puls96}, which is proportional to the luminosity to some power $x$.
        
        The key advantage when using this modified wind-momentum is that basic line driven wind theory predicts the dependence on $M_{\ast}$ to scale out (or at least to become of only second order impact). Namely, from (modified) CAK theory the following relations can be found \citep[e.g.][]{Puls08}: 
        \begin{align}\label{CAK}
            \begin{aligned}
                &\Mdot \propto L_{\ast}^{1/\alphaeff}M_{\text{eff}}^{1-\frac{1}{\alphaeff}},\\
        	    &\vinf \propto \vesc \propto \sqrt{\frac{M_{\text{eff}}}{R_{\ast}}}.
        	\end{aligned}
        \end{align} 
        Here, the effective escape speed from the stellar surface $\vesc = \sqrt{\frac{2GM_{\text{eff}}}{R_{\ast}}}$ for an effective stellar mass $M_{\text{eff}} = M_{\ast}(1-\Gamma_e)$, reduced by electron scattering
        \begin{equation}
        \Gamma_e=\frac{\kappa_e L_\ast}{GM_\ast4\pi c}
        \end{equation} 
        with an opacity $\kappa_e$ [cm$^2$/g]. Equation \eqref{CAK} above further introduces $\alphaeff = \alpha-\delta$, where $\alpha$, describing the power law distribution of the line strength of contributing spectral lines in CAK theory, takes values between 0 and 1 and the parameter $\delta$ accounts for ionising effects in the wind. For a simple $\alphaeff=2/3$ we thus have\footnote{Such a value results from considering the distribution of oscillator strengths for resonance lines within a hydrogenic ion, and neglecting $\delta$, see \citet{Puls00}.}
        \begin{align}
        	\Mdot\vinf\sqrt{R_{\ast}}\propto L_{\ast}^{1/\alphaeff},
        \end{align}
        as in the wind-momentum-luminosity relation (WLR) of equation \eqref{WMLR} for $x=1/\alphaeff$. Note, however, that even if \alphaeff\ is not exactly 2/3, the dependence on $M_{\ast}$ will still be relatively weak. The validity of the WLR relation is an overall key success of line-driven wind theory, and the basic concept has been observationally confirmed by a multitude of studies \citep[see][for an overview]{Puls08}.  
        
\begin{figure}
            \centering
            \includegraphics[width=\hsize]{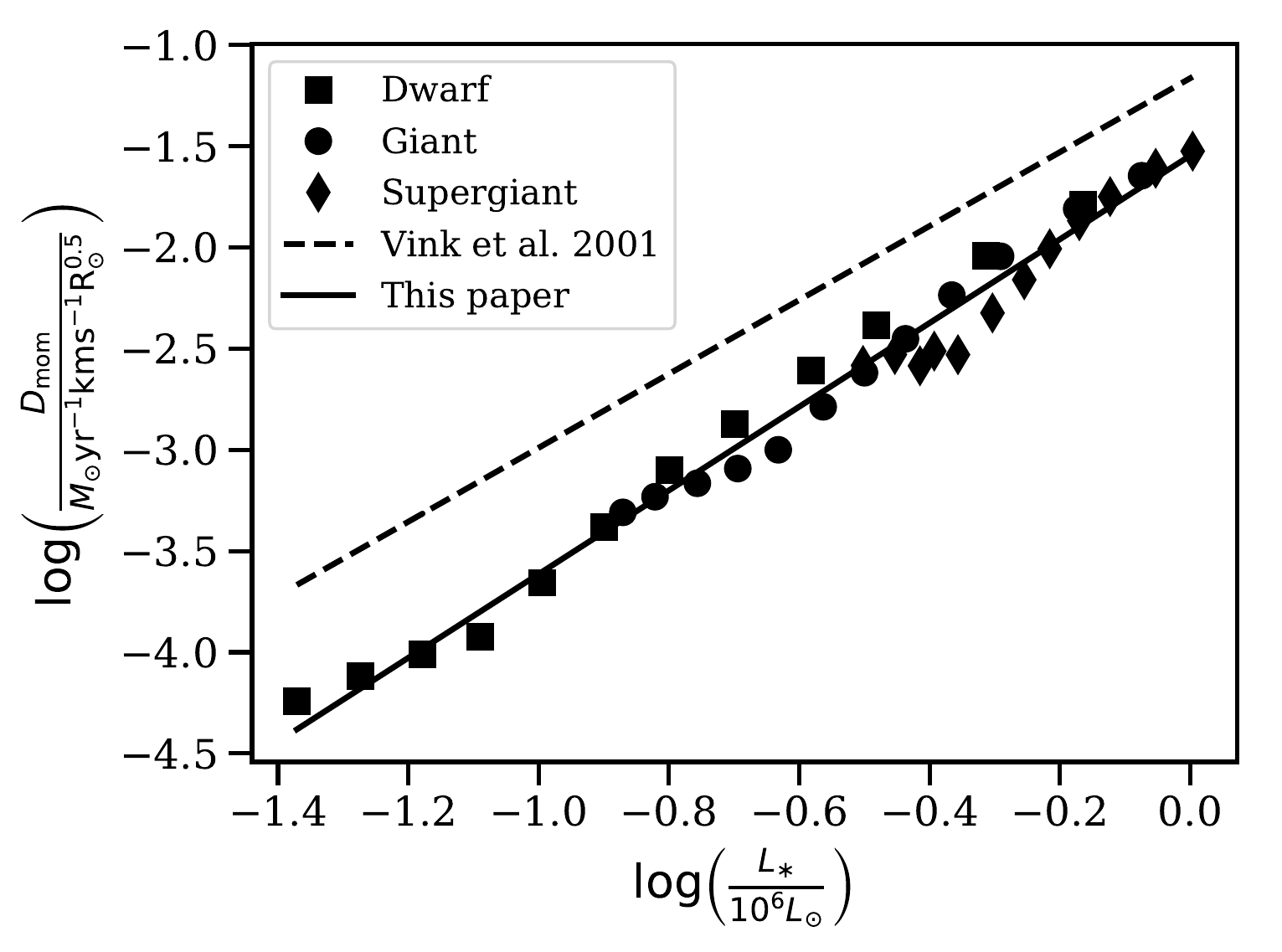}
            \includegraphics[width=\hsize]{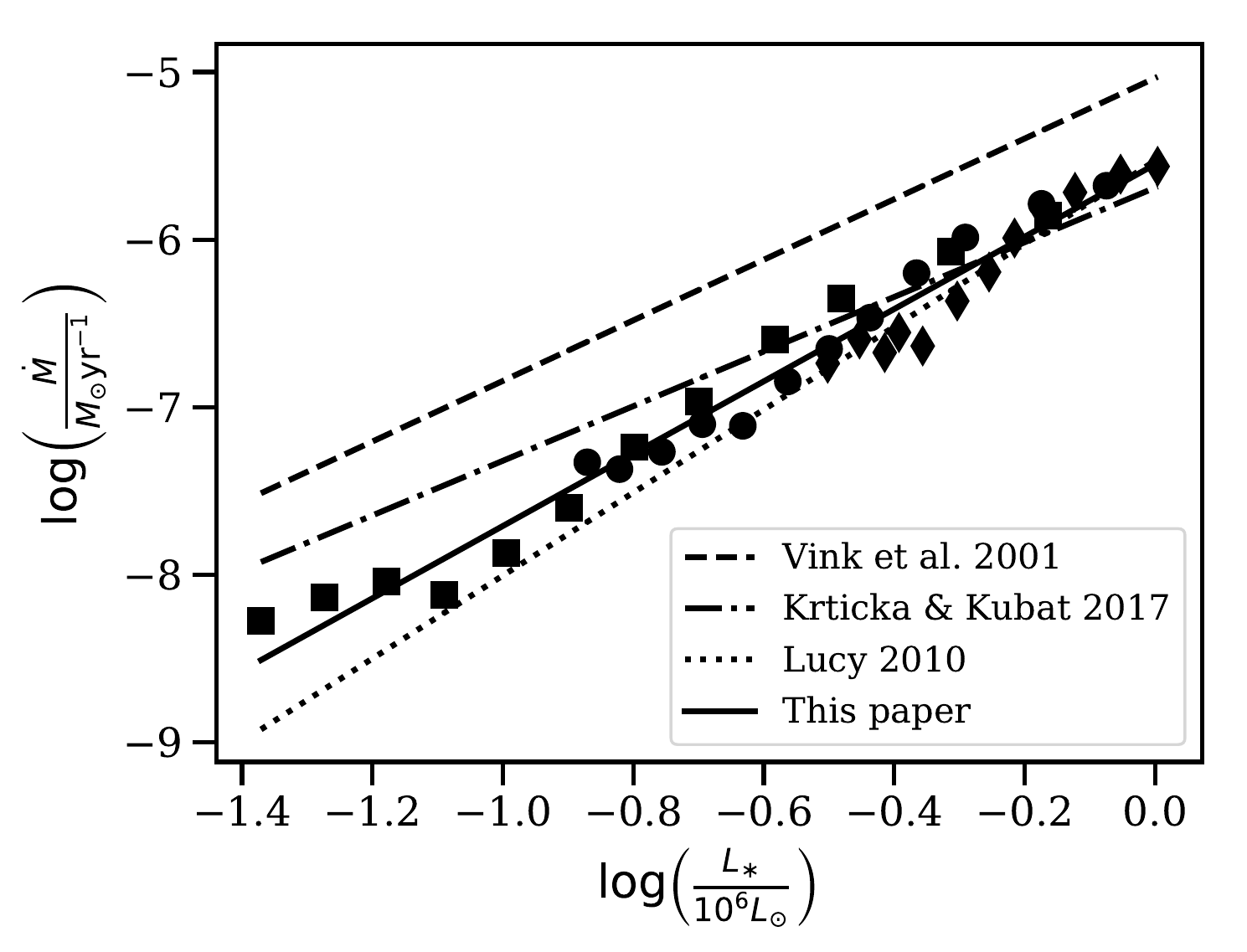}
            \caption{Top panel: Modified wind-momentum rate versus luminosity for all Galactic models. The solid black line is a linear fit through the points and the dashed line shows the theoretical relation by \cite{Vink00}. Bottom panel: Mass-loss rates versus luminosity for all Galactic models with a linear fit as a solid black line. The dashed line is a fit through the mass-loss rates computed using the Vink et al. recipe, the dash-dotted line is the relation derived by Krticka \& Kubat (2017) and the dotted line is the relation computed from the results from Lucy (2010).}
            \label{Galaxy}
        \end{figure}

        The Galactic WLR for the radiation-hydrodynamic wind models presented here is shown in the top panel of Figure \ref{Galaxy}.
        The figure indeed shows a quite tight relationship between the modified wind-momentum rate and luminosity; fitting the models according to equation \ref{WMLR} above, a slope $x = 2.07 \pm 0.32$ is derived (with the error mentioned being the 1$\sigma$ standard deviation). Interpreted in terms of the (modified) CAK theory above, this would imply a $\alphaeff\approx0.5$ for our models in the Galaxy, in rather good agreement with the typical O-star values $\alpha\approx0.6$ and $\delta\approx0.1$ \citep{Puls08}. 
        
        Next, the mass loss rate versus luminosity is plotted in the bottom panel of Figure \ref{Galaxy}. From this figure, we infer a rather steep dependence of \Mdot\ on $L_{\ast}$; fitting a simple power-law $\Mdot\propto L_{\ast}^y$ here gives $y=2.16 \pm 0.34$.
        Within our grid, we further do not find any strong systematic trends of mass-loss rate (or wind-momentum rate) with respect to spectral type (in the considered temperature range) and luminosity class. 
        
\begin{figure}
                \centering
                \includegraphics[width=\hsize]{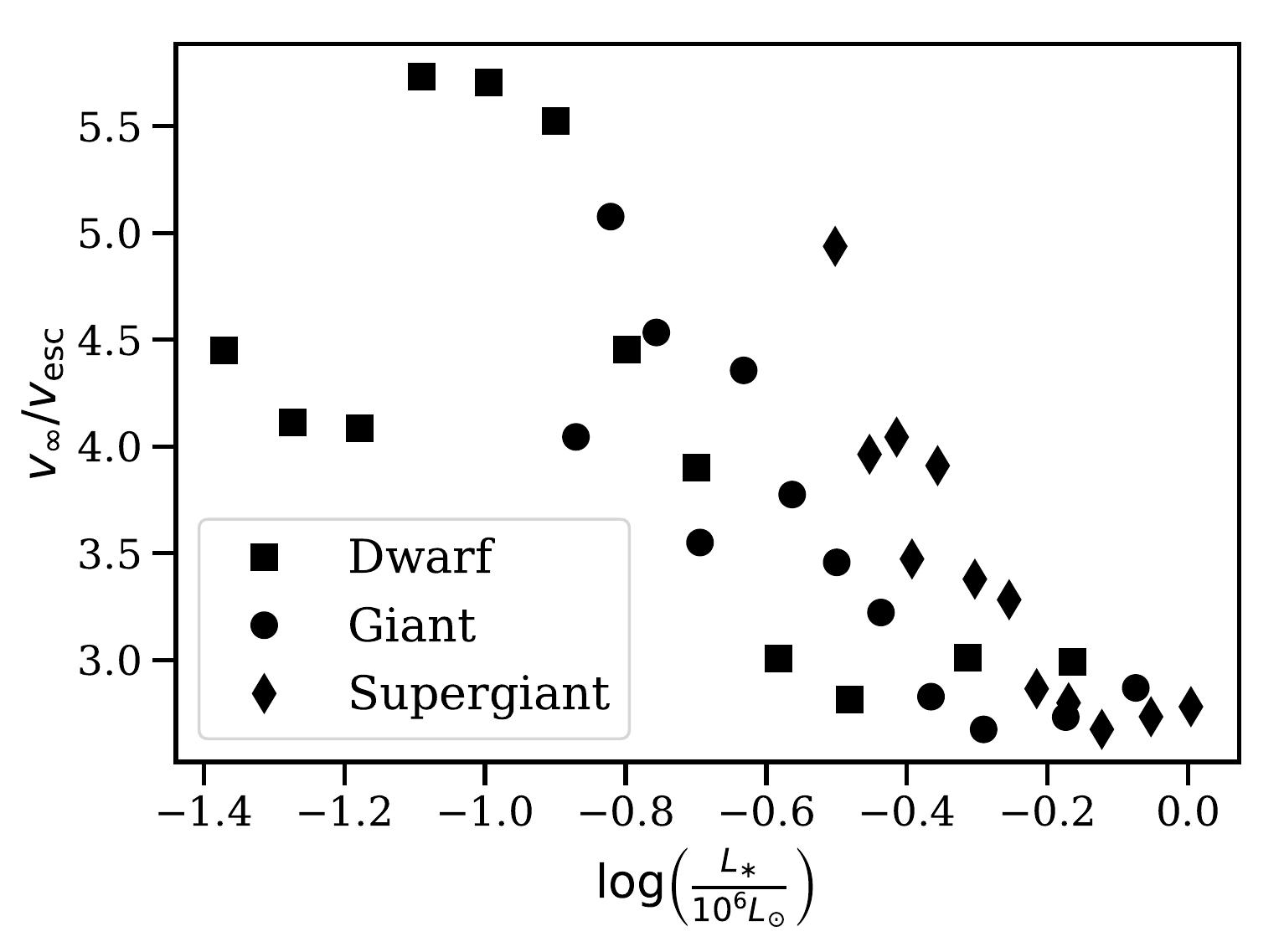}
                \caption{Terminal wind speed over photospheric effective escape speed (corrected with 1-$\Gamma_e$, see text) versus luminosity. Values for Galactic metallicity for all three luminosity classes are shown.}
                \label{vinf_vesc}
            \end{figure}

        Figure \ref{vinf_vesc} shows the terminal wind speeds for the Galactic models. We obtain a mean value $\vinf/\vesc=3.7$ for the complete Galactic sample, however with a relatively large scatter ($1\sigma$ standard deviation of 0.8). There is a systematic trend of increasing $\vinf/\vesc$ ratios for lower luminosities (see also Paper I), but also here the corresponding scatter is significant. Overall, although these Galactic $\vinf/\vesc$ values are quite high for O-stars, the significant scatter we find is generally consistent with observational studies. Sect. \ref{sec:comp} further addresses this, including a discussion about if a reduction in \vinf\ might affect also the prediction of $\dot{M}$.  
      
    \subsection{All models}
    
\begin{figure}
            \centering
            \includegraphics[width=\hsize]{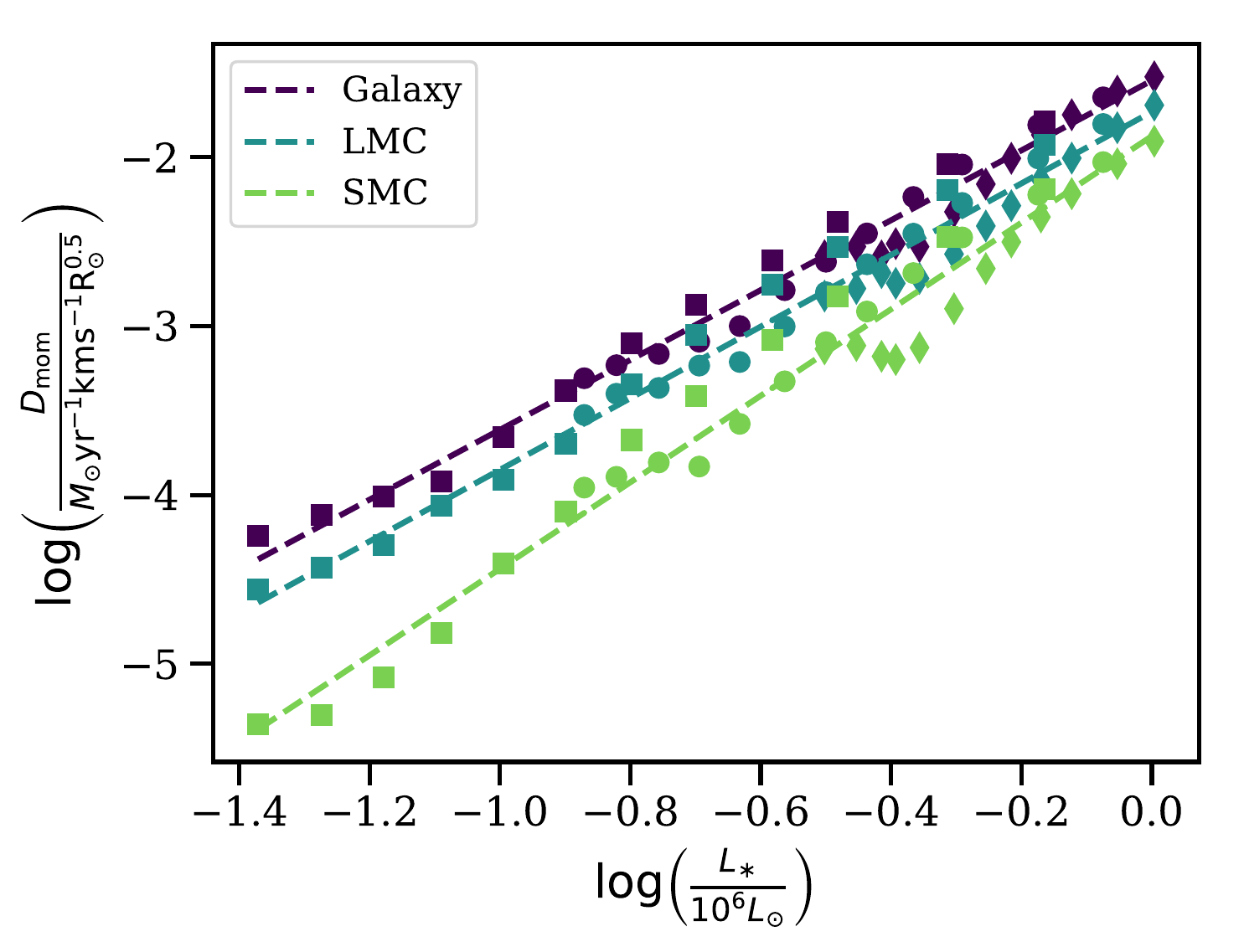}
            \includegraphics[width=\hsize]{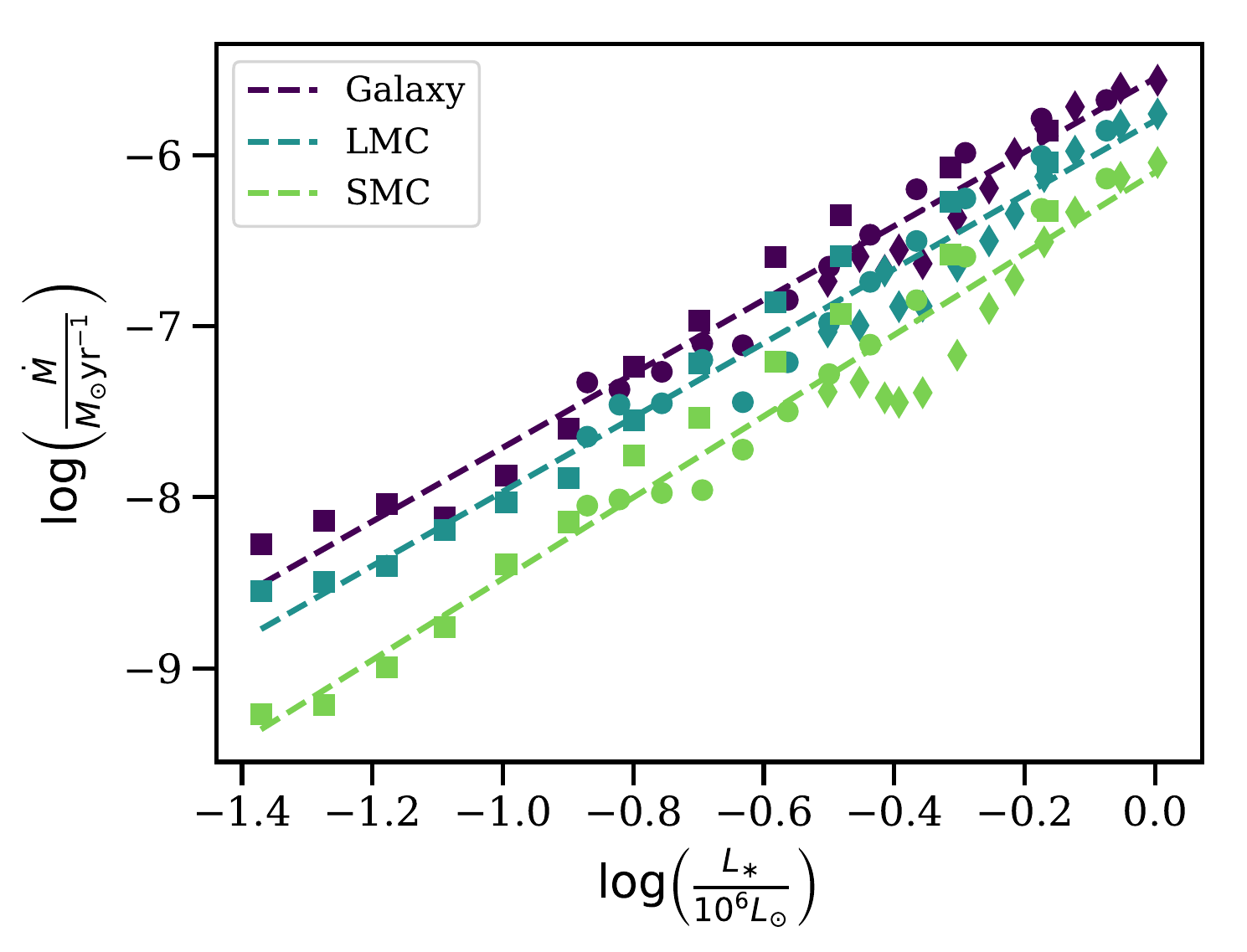}
            \caption{Top panel: Modified wind-momentum rate of all models versus luminosity. The dashed lines show linear fits through each of the three sets of models. The markers show the different luminosity classes, consistent with previous plots. Bottom panel: Same as the top panel, but for the mass-loss rate.}
            \label{all_models}
        \end{figure}

        In the top panel of Figure \ref{all_models} the modified wind-momenta for the Magellanic Cloud simulations are added to those of the Galaxy, including power-law fits also to the models in the LMC and SMC.
        The bottom panel of Figure \ref{all_models} shows the same plot for $\dot{M}$ versus $L_\ast$. On both figures there is a clear systematic trend that the mass-loss and wind-momentum rates are always lower for the LMC than for the Galaxy and lower still for the SMC. This is as expected for radiation-driven O-star winds since the majority of driving is done by metallic spectral lines. Inspection of the slope of the WLR at the LMC reveals a similar trend as for the Galaxy, and we derive $x=2.12 \pm 0.34$ from a fit according to equation \ref{WMLR}. On the other hand, already from simple visual inspection it is clear that for low-luminosity dwarfs in the SMC the overall slope changes significantly; indeed, an overall fit to equation \ref{WMLR} for all SMC stars here results in a higher $x=2.56 \pm 0.44$.
        
        Another feature visible for  the SMC supergiant models is a bump in $D_{\rm mom}$ at $\log{\frac{L_{\ast}}{10^6\Lsol}} \approx -0.3$. As discussed in section \ref{trends} this feature most likely arises because of the different effective temperatures of these models, which affects the ionisation balance of important driving elements in the wind.
        
        The derived slopes of the mass-loss rate versus luminosity, $\dot{M} \propto L_\ast^y$, for the LMC and SMC are $y=2.17 \pm 0.34$ and $y=2.37 \pm 0.40$, respectively. We note that while the dependence for the LMC is virtually unchanged with respect to the Galaxy, the SMC models again display a steeper dependence, driven by the very low mass-loss rates found for the stars with the lowest luminosities. As discussed further in Sect. \ref{sec:Discussion}, these 
        findings are generally consistent with the line-statistics predictions by \citet{Puls00} that $\alpha_{\rm eff}$ becomes lower both for lower density winds and for winds of lower metallicity.
        
    
    \subsection{Function of metallicity}\label{sec:Z}
  
\begin{figure}
            \centering
            \includegraphics[width=\hsize]{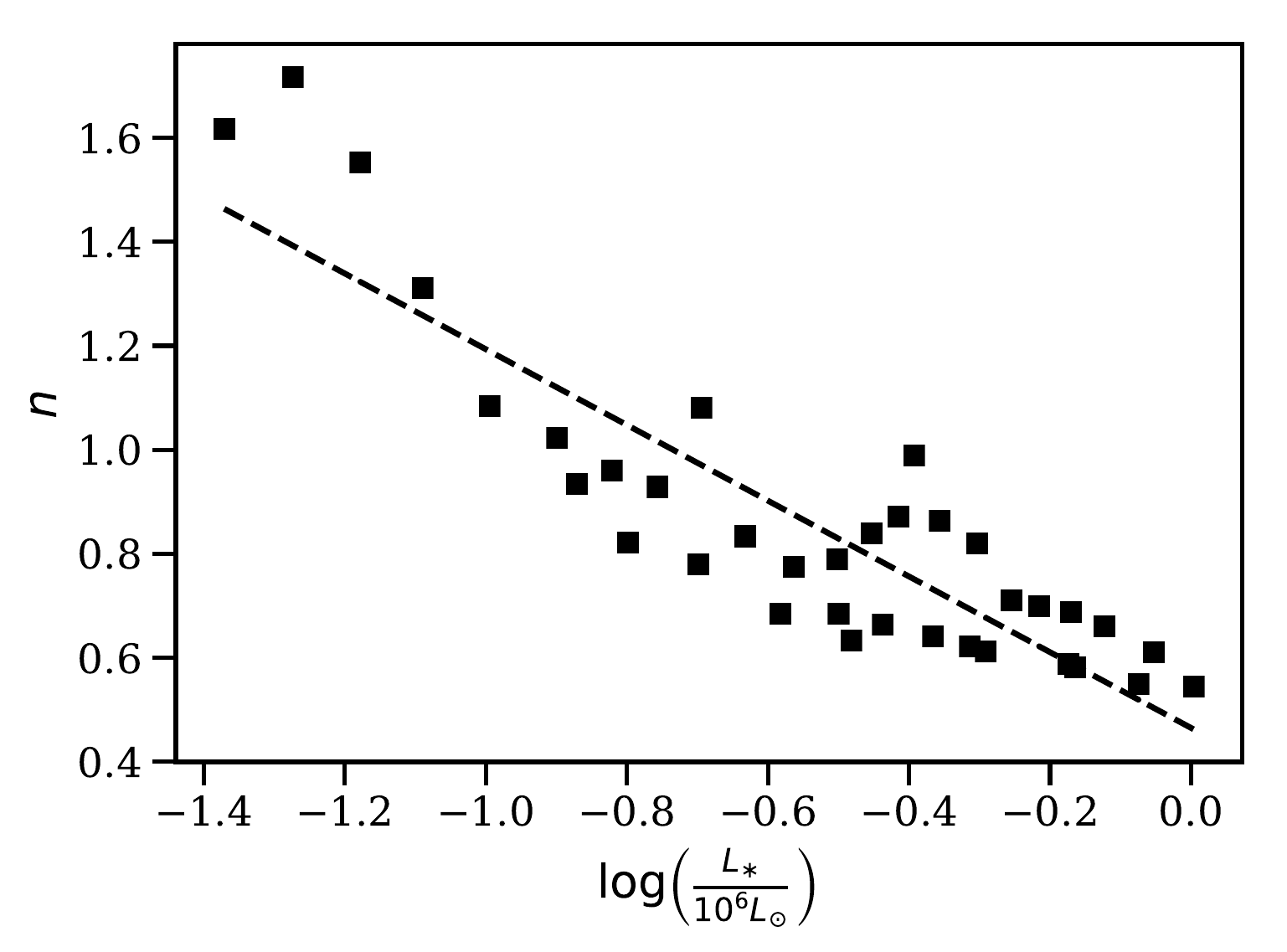}
            \caption{Value of the exponent $n$ showing the metallicity dependence of $D_{\rm mom}$ together with a linear fit, showing the linear dependence of this exponent with $\log(L_{\ast}/10^6\Lsol)$. See text for details.}
            \label{n_Mdotvinf}
        \end{figure}

        Examining trends of the modified wind-momentum rate for the Galaxy, the Large and the Small Magellanic Cloud, a dependence on metallicity is next derived. As discussed above, Figure \ref{all_models} shows that the wind-momenta of all models with the same metallicity follow a quite tight correlation with $L_\ast$. As such, to investigate the metallicity dependence we consider the three models with identical stellar parameters, but with different $Z_{\ast}$, assuming for each triplet a simple dependence 
        \begin{align}
            \Dmom \propto Z_{\ast}^n.
        \end{align}
        The derived values of $n$ are plotted in Figure \ref{n_Mdotvinf}, where the distribution gives a mean value $n=0.85$ with a 1$\sigma$ standard deviation of 0.29; this significant scatter is not  surprising since we consider only three different metallicities in the fits.
        However, inspection of Figure \ref{n_Mdotvinf} also reveals a trend of decreasing $n$ with increasing luminosity. Approximating it here by a simple linear fit with respect to $\log(L_{\ast}/10^6\Lsol)$, we find $n(L_{\ast}) = -0.73\log(L_{\ast}/10^6\Lsol) + 0.46$, providing an analytic approximation for the dependence of $D_{\rm mom}$ on $Z_{\ast}$ in function of $L_{\ast}$.
        
        The same analysis is performed also for \Mdot, assuming a dependence
        \begin{align}
            \Mdot \propto Z_{\ast}^m.
        \end{align}
        The distribution of the exponent $m$ here gives a mean value 0.95. The scatter around this mean is also significant, with a $1\sigma$ standard deviation of 0.21. If we again approximate the dependence of the factor $m$ with $L_{\ast}$, we find  $m(L_{\ast}) = -0.32\log(L_{\ast}/10^6\Lsol) + 0.79$.
        
        Building on the combined results above, we can now construct final relations for both the modified wind-momentum rate and mass-loss rate of the form: 
        \begin{align}
            \left\{
            \begin{array}{l}
                 \Dmom= A\left(\frac{L_{\ast}}{10^{6}\Lsol}\right)^y\left(\frac{Z_{\ast}}{Z_{\odot}}\right)^n,\\\\
                 \Mdot=A\left(\frac{L_{\ast}}{10^{6}\Lsol}\right)^y\left(\frac{Z_{\ast}}{Z_{\odot}}\right)^m.
            \end{array}
            \right.
        \end{align}
        To obtain the fitting coefficients (which will be different for the wind-momenta and mass loss relations), we simply combine the scalings found in Section \ref{results_G} for the Galaxy ($\Dmom\propto L_{\ast}^{2.1}$ and $\Mdot\propto L_{\ast}^{2.2}$) with those found above for the metallicity dependence ($\Dmom\propto Z_{\ast}^{n(L_{\ast})}$ and $\Mdot\propto Z_{\ast}^{m(L_{\ast})}$). Moreover, we also performed a multi-linear regression where $\log(\Mdot)$ depends on $\log(Z_{\ast})$, $\log(L_{\ast})$ and $\log(Z_{\ast})\cdot\log(L_{\ast})$; the fitting coefficients found using these two alternative methods are indeed identical to the second digit. Thus the final relations are 
        \begin{align}\label{Dmom_relation}
        \begin{split}
          \log\left(\Dmom\right) & =  -1.55 + 0.46\log\left(\frac{Z_{\ast}}{Z_{\odot}}\right) + \\
                      &\left[2.07 - 0.73\log\left(\frac{Z_{\ast}}{Z_{\odot}}\right)\right]\log\left(\frac{L_{\ast}}{10^{6}\Lsol}\right) 
        \end{split}
        \end{align}
        and
        \begin{align}\label{Mdot_relation}
        \begin{split}
          \log(\Mdot) & =  -5.55 + 0.79\log\left(\frac{Z_{\ast}}{Z_{\odot}}\right) + \\
                      &\left[2.16 - 0.32\log\left(\frac{Z_{\ast}}{Z_{\odot}}\right)\right]\log\left(\frac{L_{\ast}}{10^{6}\Lsol}\right),
        \end{split}
        \end{align}
        with the wind momentum rate in units of \Msol\ yr$^{-1}$ km s$^{-1}$ \Rsol$^{0.5}$ and the mass loss rate in units of \Msol\ yr$^{-1}$.
        When the complete model-grid sample is considered, these fitted relations give mean values that agree with the original simulations to within 10\%, with standard deviations 0.33 and 0.39 for the wind-momentum rate and the mass-loss rate, respectively. For none of the models is the ratio between the actual simulation and the fit larger than a factor 2.1 or smaller than 0.36.
        
        Finally, we can set up the same analysis to derive the dependence of the terminal wind speed on metallicity, assuming $\vinf\propto Z^{p(L_{\ast})}$. From this we find that the exponent $p$ also varies approximately linearly with the log of luminosity as $p(L_{\ast})=-0.41\log(L_{\ast}/10^6\Lsol)-0.32$. This is consistent with the results above, and indeed can be alternatively obtained by simply combining the previously derived \Mdot\ and $D_{\rm mom}$ relations (because $p=n-m$). The linear behaviour of $p$ here causes low luminosity stars to have a positive exponent with $Z$, which flattens out and gets negative for higher luminosities. So as a general trend, stars with $\log(L_{\ast}/10^6\Lsol)$ above roughly -0.78 tend to have slightly decreasing \vinf\ with increasing metallicity, while for stars below -0.78 it is the other way around.
        
            
        

\section{Discussion}\label{sec:Discussion}
    
    \subsection{General trends}\label{trends}
        The WLR results presented in the previous section show that our results would be overall consistent with standard CAK line-driven theory for $\alphaeff \approx 0.5$, at least for the Galactic and LMC cases. As already mentioned, this agrees reasonably well with various CAK and line-statistics results (see overview in \citealt{Puls08}). For the O-stars in the SMC, on the other hand, we find a steeper relation on $L_\ast$, implying an overall $\alphaeff\approx0.42$ if interpreted by means of such basic CAK theory. However, Figure \ref{all_models} shows that the WLR here exhibits significant curvature with a steeper dependence for lower luminosities, making interpretation in terms of a single slope somewhat problematic. The model-grid indicates that $\alphaeff$ decreases both with decreasing metallicity and with decreasing luminosity, suggesting that $\alphaeff$ generally becomes lower for lower wind densities. This is consistent with the line-statistics calculations by \citet{Puls00}, and may (at least qualitatively) be understood via the physical interpretation of $\alpha$ as the ratio of the line-force due to optically thick lines to the total line-force (lower wind densities should generally mean a lower proportion of optically thick contributing lines).
        
        We further find a quite steep dependence of mass loss and wind-momentum rate on metallicity.  
        Power-law fits give average values $\Dmom\propto Z_{\ast}^{0.85}$ and $\Mdot\propto Z_{\ast}^{0.95}$, however the fitting also reveals a somewhat steeper dependence for the lower-luminosity stars in our sample. The final fit-relations presented in the previous section (equations \ref{Dmom_relation}-\ref{Mdot_relation}) take this into account, and give $\Mdot\propto Z_{\ast}^{0.92}$ for the stars in our sample with luminosities above the mean ($\log(L_{\ast}/10^6\Lsol)>-0.7$) and $\Mdot\propto Z_{\ast}^{1.06}$ for the ones below this mean ($\log(L_{\ast}/10^6\Lsol)<-0.7$). Again this is generally consistent with \citet{Puls00}, who derived a scaling relation $\dot{M} \propto Z_\ast^{(1-\alpha)/\alphaeff}$. Inserting in this relation our Galactic value $\alphaeff \approx 0.48$ and a typical O-star $\delta \approx 0.1$ (see previous sections) gives $\dot{M} \propto Z_\ast^{0.9}$, whereas using the lower $\alphaeff \approx 0.42$ derived for the SMC yields $\dot{M} \propto Z_\ast^{1.1}$. These values agree rather well with the slopes we find from the scaling relations derived directly from the model-grid results. This tentatively suggests that simplified line-statistics calculations such as those in \citet{Puls00} might perhaps be used toward further calibration (and physical understanding) of the scaling relations derived in this paper, provided accurate values of $\alphaeff$ (and $\bar{Q}$, which is the line strength normalization factor due to \citealt{Gayley95}) can be extracted from the full hydrodynamic models.
        
        As for the dependence of the terminal wind speed on metallicity, the exponent seems to vary across the grid, being positive for low-luminosity stars and negative for higher luminosity stars. This might be a manifestation of two (or more) competing processes. We already found that \Mdot\ always increases when increasing metallicity, which means that winds of high $Z_{\ast}$ tend to be denser and thus harder to accelerate to high speeds. 
        On the other hand, a higher metallicity also means higher abundances of important driving lines, providing a stronger acceleration  that should increase the speed. For the low luminosity dwarfs, the second effect might be more dominant because these stars already have a low mass loss rate. As a mean value we find that \vinf\ depends on the metallicity as $\vinf\propto Z_\ast^{-0.10\pm0.18}$. Previous studies like \cite{Leitherer92} and \citet{Krticka06} find this dependence to be $\vinf\propto Z_\ast^{0.13}$ and $\vinf\propto Z_\ast^{0.06}$ respectively. While these exponents are positive, all dependencies are very shallow.
    
        As mentioned in previous sections, we do not find any strong trends with spectral type and luminosity class within our model-grid. The mass loss rates and wind-momenta follow almost the same relations for spectroscopic dwarfs, giants and supergiants. The dwarfs do show some deviation from this general trend, in particular for the SMC models, but this is mainly due to the fact that they span a much larger stellar-parameter range than the other spectral classes, thus reaching lower luminosities and so also the low mass-loss rates where the SMC WLR starts to exhibit significant curvature (see above). 
        \begin{figure}
            \centering
            \includegraphics[width=\hsize]{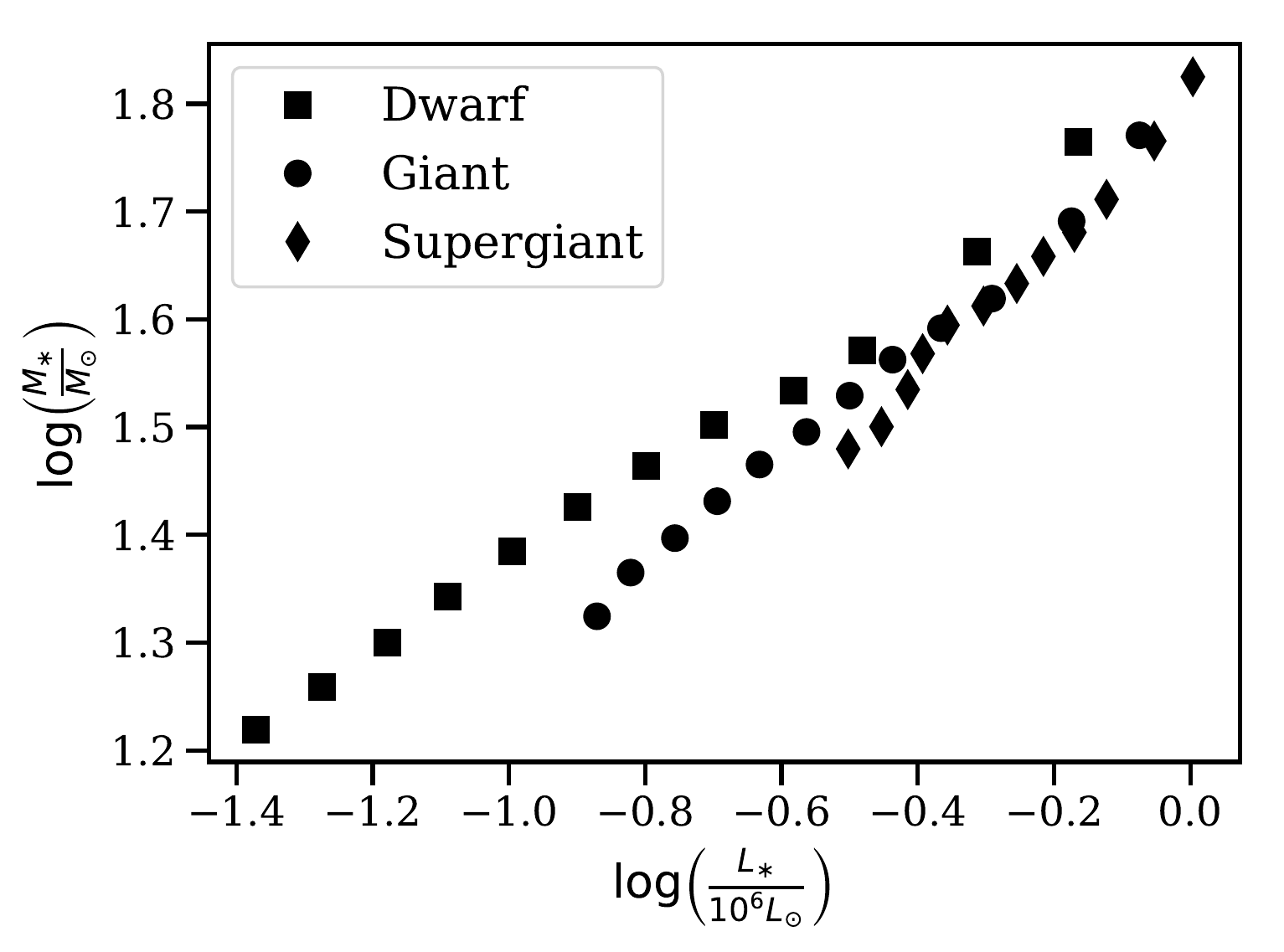}
            \caption{Mass versus luminosity \citep[from][]{Martins05} of all models, both on logarithmic scale to present a linear relation.}
            \label{ML}
        \end{figure}
        
        The dominant dependence of \Mdot\ on just $L_\ast$ and $Z_{\ast}$ within our O-star grid may seem a bit surprising, especially in view of CAK theory which predicts an additional dependence $\Mdot \propto M_{\text{eff}}^{- 0.5 \dots 1}$ (see equation \eqref{CAK}). However, including an additional mass-dependence in our power-law fits to the grid does not significantly improve the fit-quality. Moreover, a multi-regression fit to $\Mdot \propto L_\ast^a M_\ast^b$ for the Galactic sample shows that if the mass-luminosity relation within our grid $L_\ast \propto M_\ast^c$ is also accounted for, the derived individual exponents return the previously found $\Mdot \propto L_\ast^{a+b/c} = L_\ast^{2.2}$ (i.e. we get the same result as when previously neglecting any mass-dependence). Figure \ref{ML} shows this mass-luminosity relation, where a simple linear fit yields $L_\ast \propto M_\ast^{2.3}$ for our model grid. We note that these results are generally consistent with the alternative CMF models by \citet{Krticka18}, who also found no clear dependence on mass or spectral type within their O-star grid. 
        
        On the other hand, when running some additional test-models outside the range of our O-star grid, by assuming a fixed luminosity but changing the mass, we do find that including an additional mass-dependence sometimes can improve the fits. Specifically, it is clear that reducing the mass for such constant-luminosity models generally tends to increase \Mdot. This is in qualitative agreement with the results expected from CAK theory (see above), however we note that these test-models have stellar parameters that no longer fall within the O-star regime that is the focus of this paper. In any case, when extending our grid toward a more full coverage of massive stars in various phases the question regarding an explicit mass-dependency will need to be revisited.

\begin{figure}
            \centering
            \includegraphics[width=\hsize]{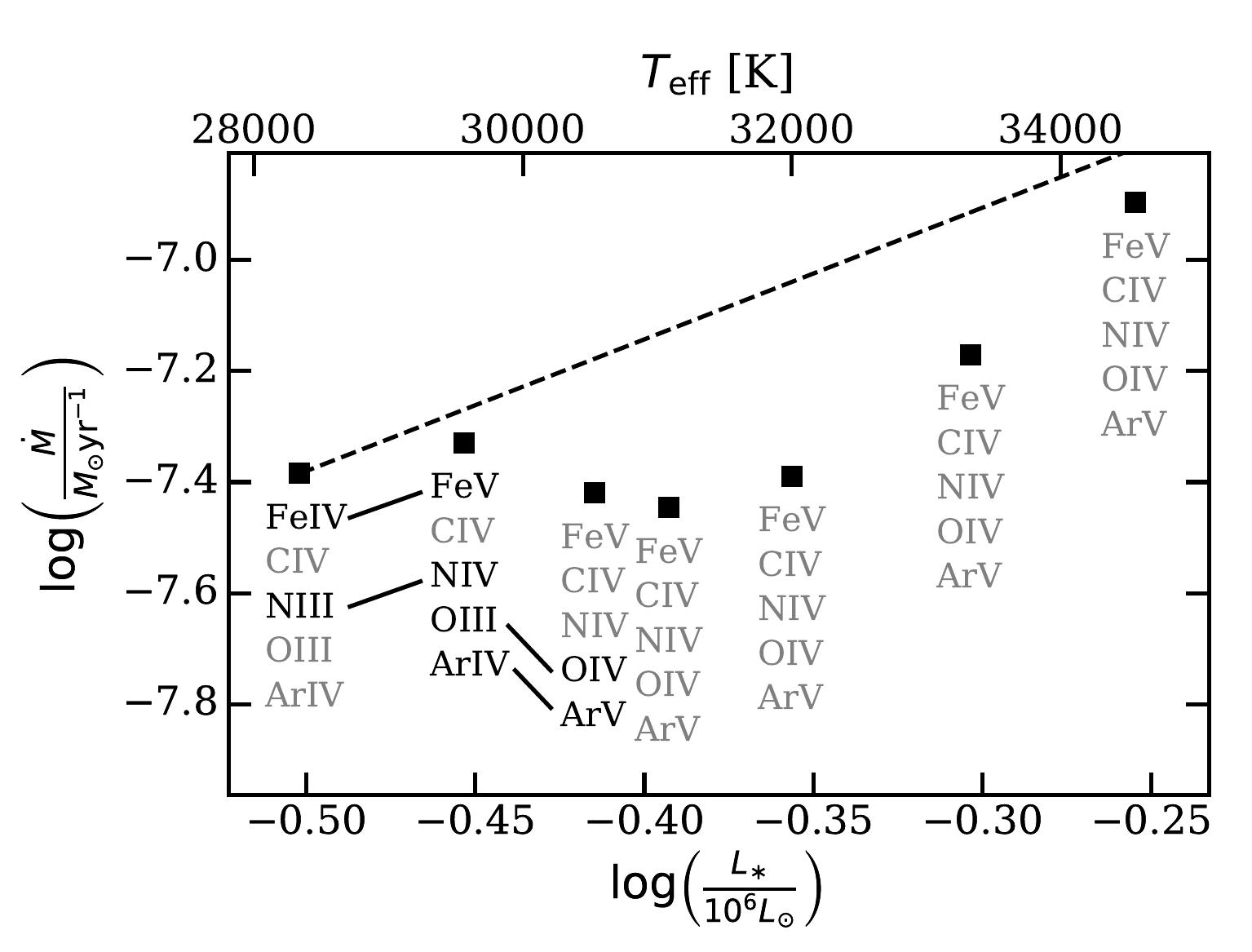}
            \caption{Zoom in on the 'bump' that is notably present for the supergiants at SMC metallicity plotted as mass loss rate versus luminosity on the lower axis and effective temperature on the upper axis. The dashed line shows the the reference slope of 2.37 derived from all SMC models. Each model shows the dominant ionisation stage at the critical point of five elements. They are connected when a transition occurs between models.}
            \label{bump}
        \end{figure}

        A notable feature for the SMC models is the bump that occurs at $\log{\frac{L_{\ast}}{10^6\Lsol}} \approx -0.3$ for the supergiants. Figure \ref{bump} shows a zoom-in on the simulations that comprise this bump. 
        For each of these models we examined the dominant ionisation stage at the sonic point of a selected list of important wind-driving elements (Fe, C, N, O and Ar). As noted above, the mass loss rate in our simulations is most sensitive to the conditions around the sonic point. If the ionisation stage of an important element changes there, the opacity and thus also the radiative acceleration may also be modified. In Figure \ref{bump} the models have increasing effective temperature and luminosity from left to right. The simple scaling relation of equation \eqref{Mdot_relation} would then predict increasing mass-loss rates from left to right, in contrast to what is observed in some of these models. As illustrated in the figure, this irregularity coincides with the temperatures at which several key driving-elements change their ionisation stages, which seem to produce a highly non-linear effect over a restricted range in the current model-grid. Although we have not attempted to explicitly account for such non-linear temperature/ionisation-effects in the fitting-relations derived in this paper, this will certainly be important to consider when extending our grids towards lower temperatures (in particular the regime where Fe IV recombines to Fe III, where this effect may become very important, e.g. Vink et al. 1999).
        
    \subsection{Comparison to other models}
\begin{figure}
            \centering
            \includegraphics[width=\hsize]{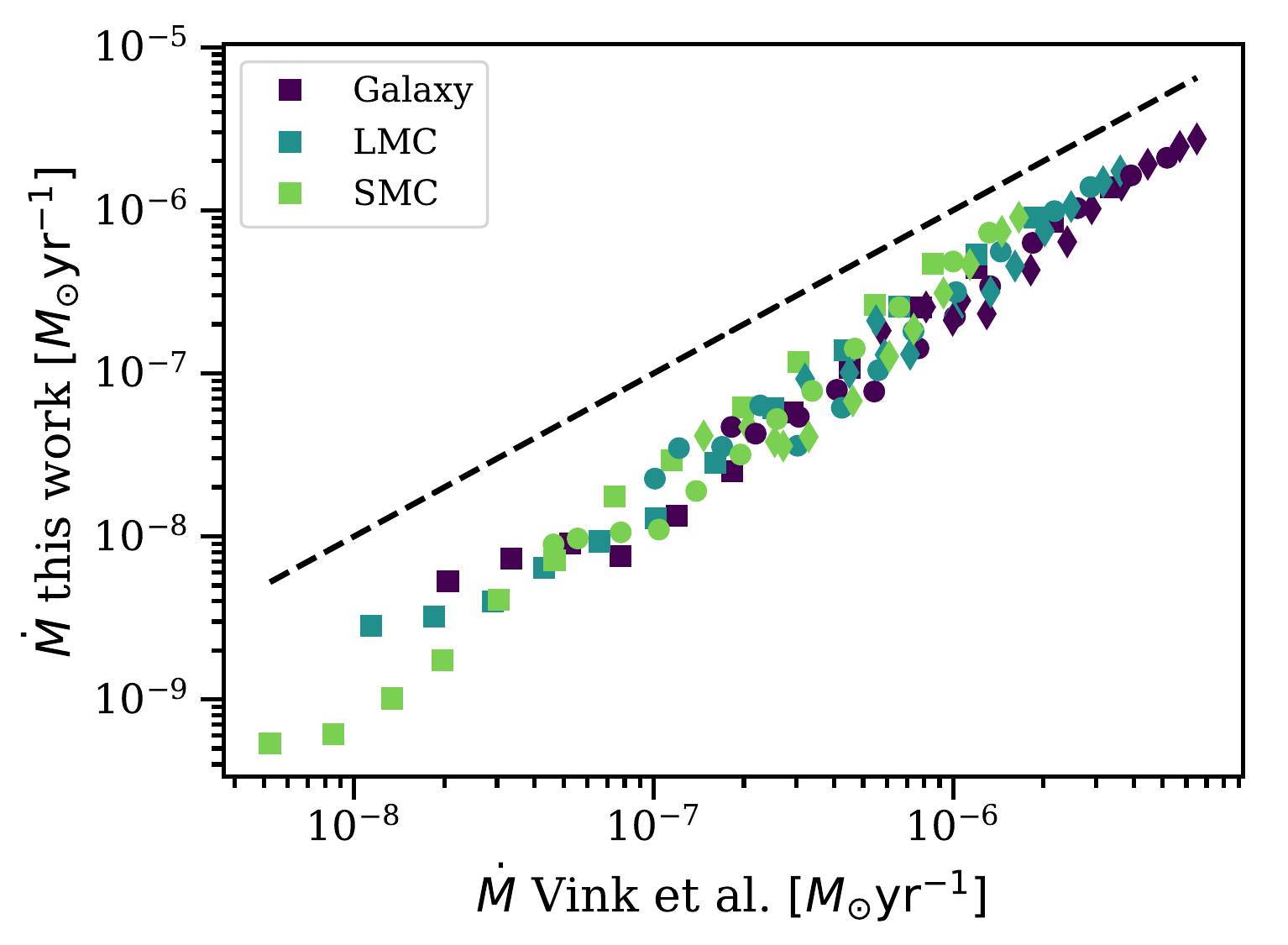}
            \caption{Direct comparison of the mass loss rates calculated in this work versus those predicted by the Vink et al. recipe. The dashed line denotes the one-to-one correspondence. The different markers show the luminosity classes consistent with previous plots.}
            \label{Vink_Leuven}
        \end{figure}

        A key result from the overall grid analysis is that the computed mass loss rates are significantly and systematically lower than those predicted by \citet{Vink00,Vink01}, which are the ones most commonly used in applications such as stellar evolution and feedback. Figure \ref{Galaxy} compares the Galactic modified wind-momenta and mass-loss rates of our models directly to these Vink et al. predictions. Their theoretical WLR presented in the top panel was constructed from a fit to the objects of their sample of observed stars. This sample was used to ensure realistic parameters for \vinf\ and $R_{\ast}$, which together with the theoretical mass loss obtained from the recipe for \Mdot\ make up the modified wind-momentum rate (see Vink et al. 2000, their Sect. 6.2). The original predictions were calculated with a higher value for metallicity of $Z_{\odot} = 0.019$, but this is scaled down here to our value of $Z_{\odot}$ to remove the metallicity effect in this comparison. Note, however, that such a simple scaling actually overestimates the effect somewhat, since the abundance of important driving elements like iron has not changed much (see Paper I). Although both set of models follow rather tight power-laws with similar exponents ($x = 2.06$ vs. $x = 1.83$), there is a very clear offset of about 0.5 dex between the two model-relations over the entire luminosity range. Indeed, every single one of our models lies significantly below the corresponding one by Vink et al. This is further illustrated in Figure \ref{Vink_Leuven}, showing a direct comparison between the mass loss rates from the full grid (i.e., for all metallicities) and those computed by means of the Vink et al. recipe.
        In addition to a systematic offset, this figure also highlights the very low mass-loss rates we find for low-luminosity stars, in particular in the SMC (reflected in the final fit-relations, equations \ref{Dmom_relation}-\ref{Mdot_relation}, through the steeper metallicity dependence at low luminosities). Regarding the metallicity scaling, the $\Mdot \propto Z_\ast^{0.85}$ obtained by \citet{Vink01} (neglecting then any additional dependence on $\vinf/\vesc$) agrees quite well with the overall values derived here. As discussed in Paper I, the systematic discrepancies between our models and those by Vink et al. are most likely related to the fact that their mass loss predictions are obtained on the basis of a global energy balance argument using Sobolev theory (and a somewhat different NLTE solution technique) to compute the radiative acceleration for a pre-specified $\beta$ velocity law. By contrast, the models presented here use CMF transfer to solve the full equation of motion and to obtain a locally consistent hydrodynamic structure (which may strongly deviate from a $\beta$-law in those regions where \Mdot\ is initiated).
        
        The bottom panel in Figure \ref{Galaxy} further compares our galactic results to those by \citet{Krticka17,Krticka18}. These authors also make use of CMF radiative transfer in the calculation of \grad, and also they find lower mass-loss rates as compared to Vink et al. However, the dependence on luminosity derived by Krticka \& Kubat is weaker than that obtained here; this occurs primarily because they predict higher mass-loss rates for stars with lower luminosities, which tends to flatten their overall slope (see Figure \ref{Galaxy}, and also Paper I). Although there are some important differences between the modelling techniques in \citet{Krticka17} and those applied here\footnote{In particular, since \citet{Krticka17,Krticka18} scale their CMF line force to the corresponding Sobolev force this means that their critical point is no longer the sonic point, and so that the nature of their basic hydrodynamic steady-state solutions may be quite different from those presented here. See also discussion in Paper I.}, the overall systematic reductions of $\dot{M}$ found both here and by them may point toward the need for a re-consideration of the mass-loss rates normally applied in simulations of the evolution of massive O-stars (as further discussed in Sect. \ref{sec:evolution} below). 
        
        A third comparison of our results, now with those from \citet{Lucy10b}, is shown in the bottom panel of Figure \ref{Galaxy}. Using a theory of moving reverse layers (MRL, \citealt{Lucy70}), they compute the mass flux $J=\rho\varv$ for a grid of O-stars with different \Teff\ and \logg. The MRL method assumes a plane parallel atmosphere and therefore does not yield a spherical mass-loss rate \Mdot\ directly. As such, we obtained \Mdot\ for the \citet{Lucy10b} simulations by simply assuming the stellar parameters of the models in our own grid, extracting the mass-loss rate from $\Mdot=4\pi R_{\ast}^2J$. The relation shown in the bottom panel of Figure \ref{Galaxy} is then computed by performing a linear fit through the resulting mass loss rates. This relation derived from the \citet{Lucy10b} models also falls well below the Vink et al. curve, even predicting somewhat lower rates than us at low luminosities. We note that in the MRL method again no Sobolev theory is used for the Monte-Carlo calculations determining $g_{\rm rad}$ and the mass flux. 
        

        
    \subsection{Comparison to observations}\label{sec:comp}
    
        Observational studies aim to obtain empirical mass-loss rates based on spectral diagnostics in a variety of wavebands, ranging from the radio domain, over the IR, optical, UV, and all the way to high-energy X-rays. As discussed in Paper I \citep[see also][for reviews]{Puls08,Sundqvist11}, a key uncertainty in such diagnostics regards the effects of a clumped stellar wind on the inferred mass-loss rate. Indeed, if neglected in the analysis such wind clumping may lead to empirical mass-loss rates that differ by very large factors for the same star, depending on which diagnostic is used to estimate this \Mdot\ \citep{Fullerton06}. More specifically, \Mdot\ inferred from diagnostics depending on the square of density (e.g., radio emission, H$_{\alpha}$) are typically overestimated if clumping is neglected in the analysis. On the other hand, if porosity in velocity space \citep[see][]{Sundqvist18b} is neglected, this may cause underestimations of rates obtained from UV line diagnostics \citep{Oskinova07,Sundqvist10, Surlan13}. Finally, regarding \Mdot\ determinations based on absorption of high-energy X-rays \citep{Cohen14}, these have been shown to be relatively insensitive to the effects of clumping and porosity for Galactic O-stars \citep{Leutenegger13,Herve13,Sundqvist18b}.
        
        \subsubsection{The Galaxy}
        
\begin{figure}
                \centering
                \includegraphics[width=\hsize]{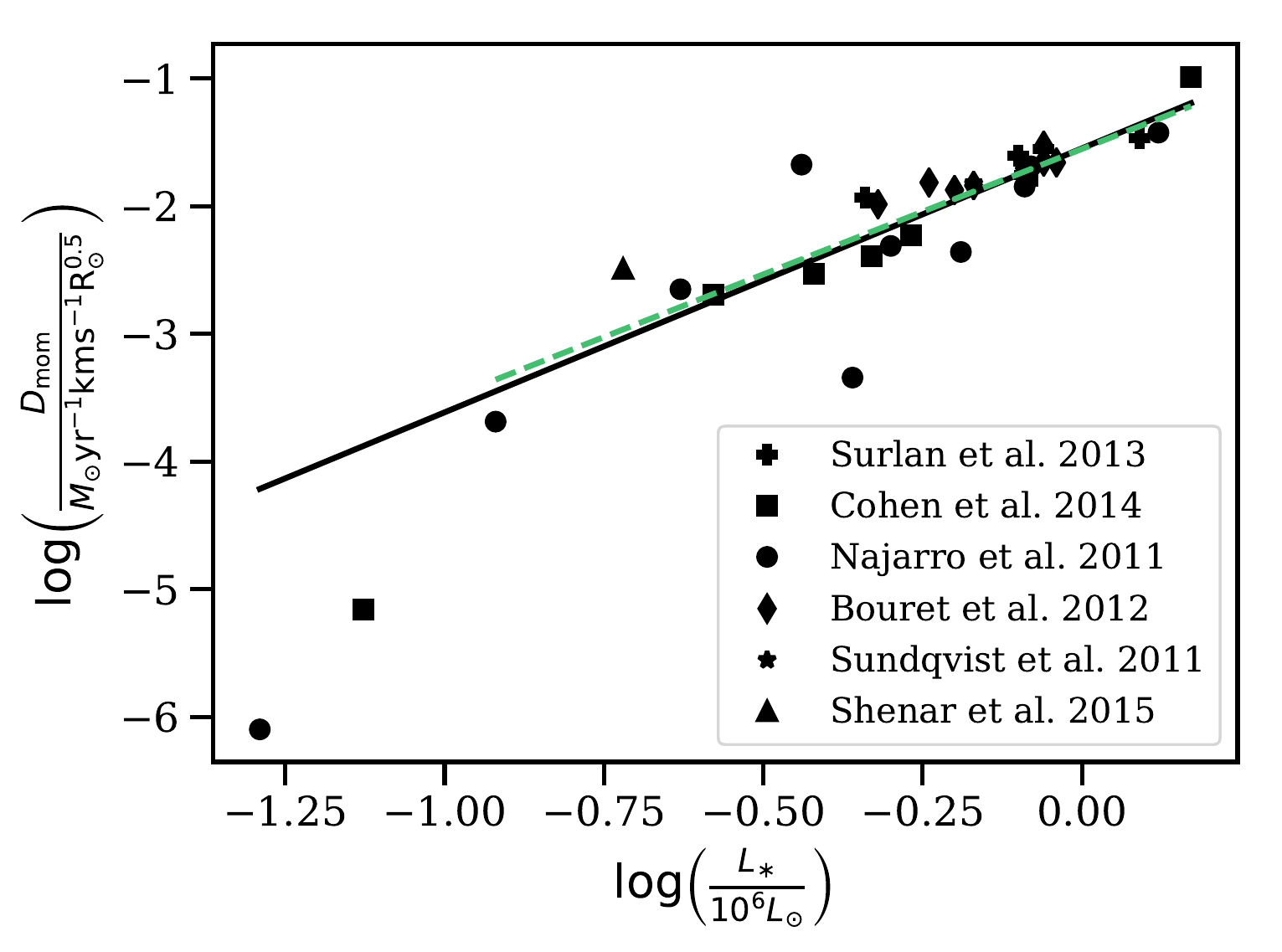}
                \caption{Observed wind-momenta for stars in the Galaxy from the studies discussed in the text are shown by different markers. The solid black line is our derived relation \ref{Dmom_relation} and the dashed line is a fit through the observations excluding the data points with $L_\ast/L_\odot < 10^5$ (see text).}
                \label{obs}
            \end{figure}

            Considering the above issues, Figure \ref{obs} compares the predictions from this paper with a selected sample of observational studies of Galactic O-star winds. The selected studies are based on X-ray diagnostics \citep{Cohen14}, UV+optical analyses accounting for the effects of velocity-space porosity \citep{Sundqvist11,Surlan13,Shenar15}, and UV+optical+IR \citep{Najarro11} and UV+optical \citep{Bouret12} studies accounting for optically thin clumping (but neglecting the effects of velocity-space porosity\footnote{Although these two last studies indeed do not account for velocity-porosity, they do attempt to adjust their studies accordingly; \citet{Najarro11} by analysis of IR lines that should be free of such effects and \citet{Bouret12} by scaling down the phosphorus abundance, thus mimicking the effect velocity-porosity would have on the formation of the unsaturated UV PV lines. As such, we opt here to include also these two studies in our selected sample for observational comparisons.}).
            The figure shows the modified wind-momentum rate from the selected observations (using different markers) together with our relation equation \eqref{Dmom_relation} (the solid line). Simple visual inspection of this figure clearly shows that our newly calculated models seem to match these observations quite well. More quantitatively, the dashed line is a fit through the observed wind-momenta, excluding the two stars with $L_\ast/L_\odot < 10^5$ but otherwise placing equal weights for all observational data points (also for those stars that are included in several of the chosen studies). For the observed stars with luminosities $L_{\ast}/L_\odot > 10^5$, there is an excellent agreement with the theoretically derived relation equation \eqref{Dmom_relation}, although there is also a significant scatter present in the data (quite naturally considering the difficulties in obtaining empirical $\dot{M}$, see above).
            
            Regarding the two excluded low-luminosity stars, these seem to indicate a weak wind effect also at Galactic metallicity. Also observational studies of Galactic dwarfs \citep{Martins05,Marcolino09} and giants \citep{deAlmeida19} find that the onset of the weak wind problem seems to lie around $\log(L_{\ast}/\Lsol) \approx 5.2$, where such an onset was already indicated in the data provided by \citet{Puls96}.
            Although our mass loss rates in this regime are significantly lower than those of \citet{Vink00}, they are still not as low as those derived from these observational studies. Further work would be required extending our current Milky Way grid to even lower luminosities, in order to examine if such an extension would yield a similar downward curvature in the WLR (albeit at a lower onset luminosity) as suggested by the observational data. Based on our results for SMC stars, we do expect that such a grid-extension might eventually start to display a significant curvature in the WLR also for Galactic objects. Nonetheless, the mismatch in the onset-luminosity between theory and observations would then still need to be explained.
            On the other hand, further work is also needed to confirm the very low empirical mass-loss rates at these low luminosities, as none of the UV-based analyses cited above account for velocity-porosity.
        
        \subsubsection{Low metallicity}
        
\begin{figure}
                \centering
                \includegraphics[width=\hsize]{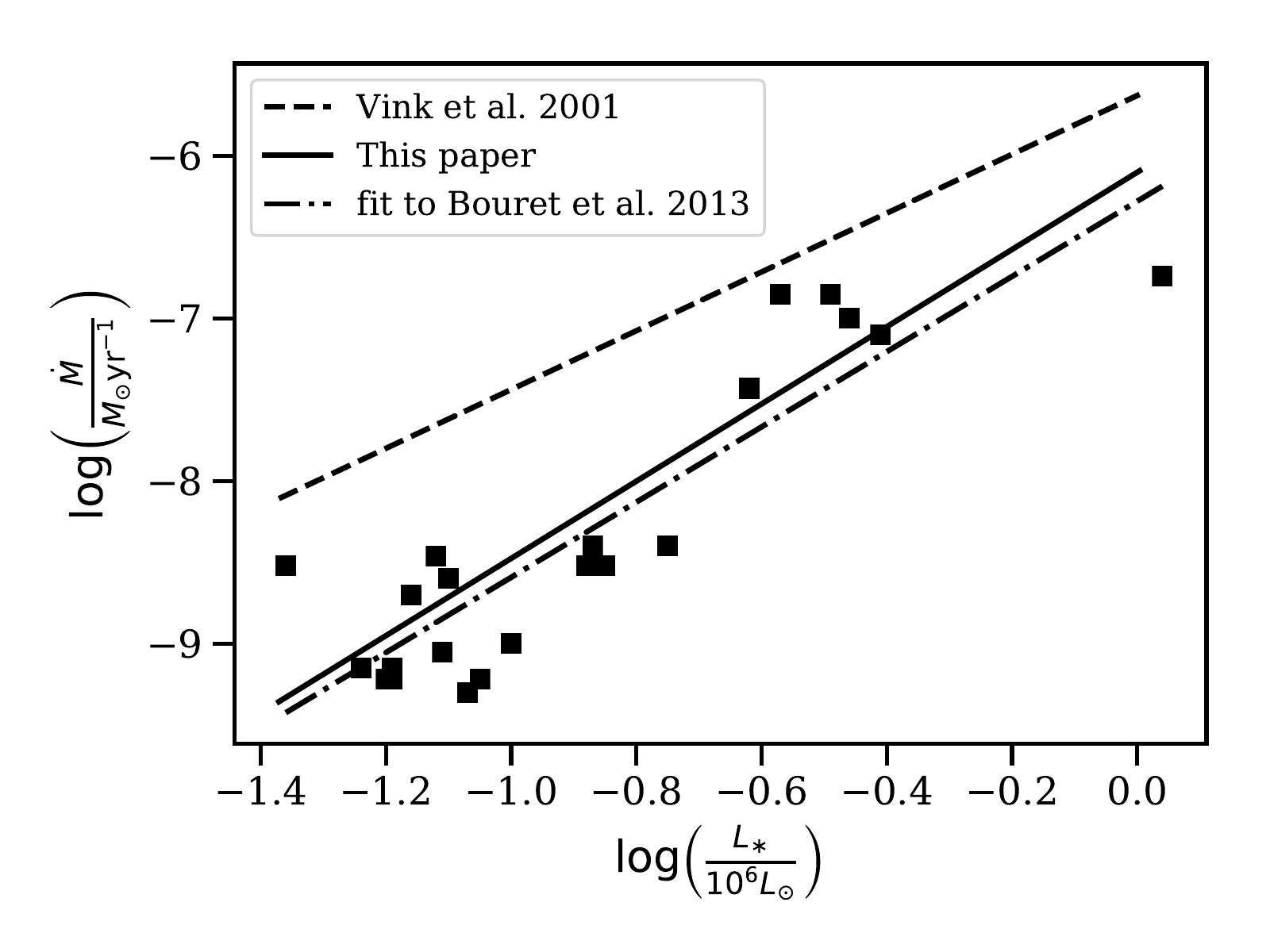}
                \caption{Observations of SMC stars from \citet{Bouret13} are shown with black squares. Both the relation as predicted by \citet{Vink01} and from equation \eqref{Mdot_relation} are shown as comparison together with a linear fit through the observational data.}
                \label{JC13}
            \end{figure}

            The above focused exclusively on Galactic O-stars, since similar analyses including adequate corrections for clumping are, to this date, scarce for Magellanic Cloud stars. 
            However, a few such studies do exist, for example the one performed by 
            \citet{Bouret13} for O dwarfs in the SMC. Again, porosity in velocity space has not been accounted for in deriving \Mdot\ from the observations in this study. Nonetheless, a comparison of our predicted rates to the empirical ones derived by Bouret et al. is shown in Figure \ref{JC13}.
           The figure displays our fit relation found for the SMC stars together with the results from \citet{Bouret13}. Additionally the figure also shows a fit to this observed sample of stars (dash-dotted line) and again also the relation for the same stars as would be predicted by the Vink et al.  recipe. It is directly clear from this figure that, again, our relation matches these observations of SMC O dwarfs significantly better than previous theoretical predictions. Indeed, since no big systematic discrepancies are found here between our predictions and the observations, 
           the previously discussed mismatch for low-luminosity Galactic O-dwarfs ("weak wind problem") does not seem to be present here for SMC conditions. This effect is also reflected through the significantly higher overall WLR slope we find for SMC models as compared to Galactic ones, $x_{SMC} = 2.56 \pm 0.44$ versus $x_{Gal} = 2.07 \pm 0.32$. In terms of line-statistics, a quite natural explanation for this behaviour is that \alphaeff\ decreases with wind density (see previous sections) which produces a steeper slope both for decreasing metallicity and decreasing luminosity. Another consequence of the found agreement could be that velocity porosity effects are negligible in the observations of SMC stars.
            
            Moreover, we may also (in a relative manner) compare our predicted scaling relations to observations in different galactic environments, as long as we assume that clumping properties do not vary significantly between the considered galaxies (and so does not significantly affect the relative observational results).
            In a large compilation-study of O-stars in the Galaxy, LMC, and SMC, \citet{Mokiem07b} obtained observationally inferred values of $D_{\text{mom}}$. Specifically, \citet{Mokiem07b} use the H$_{\alpha}$ emission line to derive the mass loss rate. This diagnostic is highly dependent on the clumping factor, but comparison with our scaling results will still be reasonable if the clumping in the H$_{\alpha}$ forming region does not vary too much between the different metallicity environments. Neglecting such potential clumping effects, Mokiem et al. derived the empirical WLR slopes $x_{Gal} = 1.86 \pm 0.2$, $x_{LMC} = 1.87 \pm 0.19$, and $x_{SMC} = 2.00 \pm 0.27$. Within the 1$\sigma$ errors, this agrees with the theoretical values $x_{Gal} = 2.07 \pm 0.32$, $x_{LMC} = 2.12 \pm 0.34$, and $x_{SMC} = 2.56 \pm 0.44$ obtained here (although the general trend is that we find somewhat steeper relations). In this respect, we note also that the tight 1$\sigma$ limits in the empirical relations obtained by \citet{Mokiem07b} only reflected their fit-errors, and not additional systematic errors due to uncertainties in, e.g., stellar luminosities and adopted metallicities. As such, the overall agreement between their empirical relations and the theoretical WLR slopes presented in this paper is encouraging. Moreover, by considering their results at a fixed luminosity $\log L_\ast/L_\odot = 5.75$, \citet{Mokiem07b} also derived an empirical mass-loss vs. metallicity relation $\dot{M} \propto Z_{\ast}^{0.83 \pm 0.13}$; again this agrees rather well with the theoretical $\dot{M} \propto Z_{\ast}^{0.87}$ obtained at such a $\log L_\ast/L_\odot = 5.75$ from equation \eqref{Mdot_relation} of this paper.

        \subsubsection{Terminal wind speed}
        
            As can be seen in Figure \ref{vinf_vesc}, $\vinf/\vesc$ ratios range between approximately 2.5 to 5.5 for our Galactic models. Similar values are found for the Magellanic Cloud simulations, but the scatter is significant and it is difficult to draw general conclusions from the model data. A significant scatter in $\varv_\infty/\varv_{\rm esc}$ is however quite consistent with observational compilations such as the one presented by \citet{Garcia14} (see in particular their Fig. 9) and \citet{Lamers95}. Overall though, our mean values are somewhat higher than observations generally suggest (see also \citealt{Kudritzki00}), in particular for low-luminosity objects in the Galaxy. This issue was also addressed in Paper I, where it was argued that the high $\varv_\infty$ for such low-density winds might be naturally reduced as a consequence of inefficient cooling of shocks in the supersonic wind (see also \citealt{Lucy12}). Namely, if a significant portion of the supersonic wind were shock-heated this would lower $g_{\rm rad}$ in these layers and so also potentially reduce $\varv_\infty$. Moreover, if the wind is shock-heated, empirical estimates of $\varv_\infty$ might be misinterpreted as it may no longer be possible to accurately derive this quantity from observations of UV wind lines. Tailored radiation-hydrodynamical simulations of line-driven instability (LDI) induced shocks in low-density winds are underway in order to examine this in detail, and will be presented in an upcoming paper (Lagae et al., in prep.). For now we study how a reduced wind driving in supersonic layers may affect $\varv_\infty$ and $\dot{M}$ in our steady-state simulations, by running a few additional models where $g_{\rm rad}$ in layers with speeds above 1000 km/s simply is reduced by an ad-hoc factor of two. While these models then indeed converged to a significantly lower $\varv_\infty$ than previously, $\dot{M}$ was only marginally affected. This suggests that (at least within the steady-state simulation set-up used here) a reduction of the outer wind driving does not necessarily lead to an increased mass-loss rate, because in our simulations \Mdot\ is mostly sensitive to the conditions at the sonic point and is not as much affected by the reduction of \grad\ in the highly supersonic regions.
            Although shock-heating in the outer winds in low-luminosity objects might explain the high predicted values of the low luminosity objects, it is unclear to what degree the increasing trend in \vinf/\vesc\ for decreasing $L_{\ast}$ might still persist in such models. Similarly, it is also somewhat unclear how this matches with observations, as \citet{Lamers95} do not find such a trend while \citet{Garcia14} do show a similar trend, at least for dwarf stars. All these observations show significant scatter though, such that a clear prediction is difficult to obtain from the present data. To a major part, the scatter in \vinf\ is understandable though, since in the outer wind only a few dozen strong resonance lines are responsible for the acceleration (mostly from C, N, O, Ne and Ar). Thus, with only a few lines, a certain difference in the abundances (e.g., because of different ages and mixing) can have a significant effect on \vinf, leading to the observed scatter.

        \subsection{Influence of $\varv_{\rm turb}$}
        
        As discussed in Paper I, the turbulent velocity \vturb\ can have a significant effect on $g_{\rm rad}$ around the sonic point, and so also affect the predicted \Mdot\ . In general, increasing \vturb\ tends to decrease \Mdot\ (\citealt{Lucy10b}; \citealt{Krticka17}; Paper I) mostly because higher velocities then are required to Doppler shift line profiles out of their own absorption shadows, reducing $g_{\rm rad}$ in the critical layers. To study the behaviour of our grid with \vturb\, we ran a set of additional simulations where, for each of our previous models, we increased \vturb\ to 12.5 km/s and decreased it to 7.5 km/s. This way a slope of $\log\Mdot$ with $\log\vturb$ could be found for each model in the grid. The mean value obtained is  
        \begin{align}\label{vt_ours}
            \frac{\partial\log\Mdot}{\partial\log\vturb} = -1.06\pm0.40,
        \end{align}
        where the error is derived from the $1\sigma$ spread of the slope for all models. Also \citet{Lucy10b} studied the behaviour of the mass flux $J$ with \vturb. Like here he finds an inverse dependence, but with a somewhat steeper slope -1.46. Note, however, that this slope was derived from a single model at \Teff\ = 40000 K and \logg\ = 3.75 whereas we here consider an average across our full grid. Indeed, inspection of a single model similar to the one used in \citet{Lucy10b}, with \Teff\ = 40062 K and \logg\ = 3.92, reveals a slope -1.49 that is in good agreement with their result. Furthermore, considering the large scatter on the slope that we find, with values between a maximum of -0.07 and a minimum of -1.88, the result of equation \eqref{vt_ours} is also still in agreement with that of \citet{Lucy10b}.
        But although the scatter around our derived mean-slope thus is significant, there are no clear trends within the grid. As such, to obtain a first-order approximation accounting for a \vturb\ that deviates from the standard value 10 km/s, the  \Mdot\ relations in Sect. 3.3 may simply be scaled according to equation \eqref{vt_ours}.
        
        \subsection{Comparison to $\beta$ velocity law}\label{beta}
        
        Most wind models of hot, massive stars that are used for spectroscopic studies actually do not solve the e.o.m. for the outflow. Instead, these models assume an empirical "$\beta$-type" wind velocity-law that connects to a quasi-hydrostatic photosphere. This is the case also for the "standard" version of {\sc fastwind}, used here as a starting condition for our self-consistent simulations (see Sect. 2). For the prototypical simulation presented in Section \ref{Model_outcome}, Fig \ref{vr} illustrates a fit to the self-consistent calculated velocity using such a $\beta$-law. More specifically, a "double" $\beta$-law  similar to that presented in Sect. 5.2 of Paper I is used, however in order to obtain a better fit to the very steep acceleration of the transition region we have here also included a modification-term in dependence on the photospheric scale-height.
        The full expression used to match the velocity structure is
        \begin{align}\label{2beta}
            \varv(r) = \frac{(\vinf-\varv_{\rm exp})\left(1-\frac{r_{\rm tr}}{r}\right)^{\beta}+\varv_{\rm exp}}{1+\left(\frac{\varv_{\rm exp}}{\varv_{\rm tr}}-1\right)\exp\left(\frac{r_{\rm tr}-r}{H}\right)},
        \end{align}
        where
        \begin{align}
            \beta = \beta_1+(\beta_2-\beta_1)\left(1-\frac{r_{\rm tr}}{r}\right).
        \end{align}
        In this relation, the parameters \vinf, $\beta_1$, $\beta_2$, $\varv_{\rm exp}$ and $H$ are obtained by fitting to the numerically derived velocity structure above the transition radius $r_{\rm tr}$, defined according to $\varv(r_{\rm tr}) = \varv_{\rm tr}$. Introduced here is the parameter $\varv_{\rm exp}$, which is roughly the velocity at which $\varv(r)$ has its biggest curvature and thus controls how far the exponential behaviour of the velocity holds in the inner wind. Also introduced is $H$, which is the scale height setting the stratification in the photosphere, controlled by the density structure of the model.
        
        With this the model displayed in Fig. 4 can be well fit down to a transition velocity $\varv_{\rm tr} \approx 0.1 a$, for a primary beta-factor $\beta_1 = 0.8$. We note though, that in particular the more low-density winds in our sample typically require much higher transition velocities in order to be well fit, often values well above $\varv_{\rm tr} \approx 0.5a$ are found (see also discussion in Paper I). Moreover, most of the best-fit $\beta_1$-values lie between 0.5-0.8, indicating a steep acceleration of the inner wind across our grid. A more detailed study aiming to develop an improved  velocity-parameterization for spectroscopic studies is underway, and will be presented in an upcoming paper. There, we also plan to examine in detail what effects the predicted steep  acceleration in the transition region may have upon the formation of various strategic spectral lines used for diagnostic work.    

    \subsection{Implications for stellar evolution}\label{sec:evolution}
    
        The stellar mass is the most important parameter defining the evolution of a star, and accurate mass loss rates are crucial to determine the corresponding evolutionary pathways. Codes for stellar structure and evolution use prescribed recipes calculating the change in stellar mass in between time steps. Many codes, like MESA \cite[][and references therein]{Paxton19}, use the Vink et al. recipe to calculate \Mdot\ for hot, hydrogen-rich stars (excluding then the classical Wolf-Rayet (WR) stars which require different prescriptions for \Mdot). As clearly illustrated by Figure \ref{Vink_Leuven}, the results presented here suggest that the O-star mass-loss rates should be significantly lower. In addition, this is also supported by the comparison to observations in Section \ref{sec:comp}. Even though the calculations here are performed for massive stars in their early phases of evolution (mostly on the main-sequence), the lower rates will not only affect the stars there, but also impact the properties of post-main-sequence stages. One consequence is that the luminosity at which the stars end their main-sequence evolution and cross the Hertzprung gap is changed, as seen directly in our evolution calculations of a 60\Msol\ star using \MESA\ (where we have simply reduced the amount of mass loss on the main sequence to be in accordance with the models presented here). Another effect is that a lower mass loss means that angular momentum is lost less rapidly so that the star keeps a higher surface rotation. \citet{Keszthelyi17} computed models with reduced mass-loss rates (factor 2 to 3) and found the surface rotation speeds at the end of the main-sequence to remain rather high, possibly requiring an additional source of angular momentum loss to reconcile with observed values.
        
        Moreover, \citet{Belczynski10} studied the masses of compact objects originating from single stars. Here, the adopted mass loss during the life of the progenitor star is crucial in determining the maximum black hole mass that can be achieved. The determinations of the black hole masses in \citet{Belczynski10} were done assuming the Vink et al rates; adopting lower rates would reduce the amount of mass lost and thus possibly increase the resulting black hole mass. Getting direct measurements of these black hole masses is not straightforward. However, by taking advantage of gravitational wave astronomy, detections like GW150914 \citep{Abbott16} provide observational constraints. 
        Derived black-hole masses turn out to be relatively high, $\gtrapprox 25 \Msol$, which, with current wind prescriptions, can only be created in low metallicity environments. 
        With lower values of the mass-loss rate during the evolution of the star, such as proposed here in this paper (or like in magnetic massive stars, \citealt{Petit17}), the "heavy" black holes as detected by gravitational waves might in principle be produced also in high-metallicity environments \citep{Belczynski20}. 
        So far though, in the Galaxy, no observations of such heavy mass black holes (from single stars) exist as all of them have a mass less than 15\Msol\ \citep[as found in studies like][]{Shaposhnikov07,Torres20}. Depending on the initial mass of the progenitor stars, these values can be explained both by our proposed mass loss rates as well as with those predicted by \cite{Vink01}.
        
        Significant mass loss is further necessary to create the naked Helium core that is a classical WR-star through wind-stripping. In low metallicity environments such as the SMC, this is generally considered to be difficult because of the strong metallicity dependence resulting in low mass-loss rates. Considering our results here of lower O-star mass-loss rates, such (steady) wind-stripping would be more difficult to obtain also in the Galaxy, potentially leading to an increase of the lower limit for the initial mass of WR-stars created by this channel. In order to explain the observed number of WR-stars, alternative pathways might thus be necessary. In this respect, a straightforward option is binarity where the outer layers of stars can be removed through binary interaction like Roche-lobe overflow \citep[e.g.][]{Gotberg18}, and recent studies have shown that a large majority of massive stars resides in such binary systems \citep{Sana14}. On the other hand, a second possible channel is eruptive mass loss in the luminous blue variable stage (LBV) \citep{Smith14}. Indeed, significant fractions of the stellar mass can be removed in such eruptive events; the LBV $\eta$-Carina, for example, lost $10\Msol$ in just 10 years in the 19th century. Considering also that the (so far confirmed) binary fraction of WR-stars in the SMC is lower than that of the Galaxy, or at least similar \citep{Foellmi03}, this latter pathway might prove to be of increased importance.
        

\section{Summary and future prospects}\label{sec:Conclusions}

    We calculated a grid of steady-state wind models of O-stars by varying fundamental stellar parameters in three metallicity regimes corresponding to the Galaxy, the Small and the Large Magellanic Clouds. The models provide predictions of global wind parameters like mass-loss rate and wind-momentum rate, allowing us to analyze how these quantities depend on fundamental stellar parameters like luminosity and metallicity.
    
    From our grid we find steep dependencies of the mass-loss rate with both luminosity and metallicity, mean values $\Mdot\sim L_{\ast}^{2.2}$ and $\Mdot\sim Z_{\ast}^{0.95}$.
    The metallicity dependence is further found to vary across the luminosity range, and accounting for this results in the final fit relations for the wind-momentum rate and mass loss rate presented in Section \ref{sec:Z}. Additionally, a clear change in slope for the predicted WLR for dwarfs in the SMC is found, pointing toward the occurrence of weak winds in the models.
    
    Our computed mass-loss rates are significantly lower for all models than those predicted by Vink et al. (2000,2001), which are the ones usually implemented in evolution calculations of massive stars. Such lower O-star wind-momenta and mass-loss rates are also in general 
    accordance with observational studies in the Galaxy that account properly for the effects of clumping upon the diagnostics used to infer the empirical mass-loss rates. Regarding the metallicity dependence, our scaling predictions are (within the errors) in agreement with the larger empirical study by \citet{Mokiem07b}.
    
    The systematically reduced mass-loss rates for all models strengthens the claim that new rates might be needed in evolution simulations of massive stars. Namely, adopting different rates can significantly affect the evolution of the massive star, for example by changing its spin-down time and altering the initial mass needed in order to produce a wind-stripped Wolf-Rayet star. As such, a key follow-up work to this study will be to now extend the grid presented here to include massive stars outside the O-star domain, to incorporate our new models into simulations of massive-star evolution, and to analyze in detail the corresponding effects.
    
\begin{acknowledgements}
      RB and JOS acknowledge support from the Odysseus program of the Belgian Research Foundation Flanders (FWO) under grant G0H9218N.
      JOS also acknowledges support from the KU Leuven C1 grant MAESTRO C16/17/007
      FN acknowledges financial support through Spanish grants ESP2017-86582-C4-1-R and PID2019-105552RB-C41 (MINECO/MCIU/AEI/FEDER) and from the Spanish State Research Agency (AEI) through the Unidad de Excelencia “María de Maeztu”-Centro de Astrobiología (CSIC-INTA) project No. MDM-2017-0737.
      We would also like to thank the referee for useful comments that led to additional improvements of the paper.
\end{acknowledgements}
    
\bibliographystyle{aa} 
\bibliography{all_papers.bib} 

\appendix
\section{Model parameters}\label{appendix}

    \onecolumn
    \begin{longtable}{ c c c c c c c c }
        \caption{Input parameters of all models in the grid together with the resulting mass-loss rate and terminal wind speed.}\label{all_param}\\
        \hline\hline
        $\log(L_{\ast}/L_{\odot})$ &	$M$ [$\Msol$] &	$R_{\ast}$ [$\Rsol$] & \Teff\ [K] &	$Z_{\ast}$ [$Z_{\odot}$] & $\Gamma_{\rm{e}}$ &	$\log(\Mdot)$ [$\Msol$/yr] &	$\vinf$ [km/s]\\ \hline
        \endfirsthead
        \caption{continued.}\\
        \hline\hline
        $\log(L_{\ast}/L_{\odot})$ &	$M$ [$\Msol$] &	$R_{\ast}$ [$\Rsol$] & \Teff\ [K] &	$Z_{\ast}$ [$Z_{\odot}$] & $\Gamma_{\rm{e}}$ &	$\log(\Mdot)$ [$\Msol$/yr] &	$\vinf$ [km/s]\\ \hline
        \endhead
        \hline
        \endfoot
	$            5.50$ &	$           30.18$ &	$           23.11$ &	$           28430$ &	$             0.2$ &	$            0.27$ &	$         -7.3839$ &	$         3686.28$\\
	$            5.50$ &	$           30.18$ &	$           23.11$ &	$           28430$ &	$             0.5$ &	$            0.27$ &	$         -7.0341$ &	$         3387.06$\\
	$            5.50$ &	$           30.18$ &	$           23.11$ &	$           28430$ &	$             1.0$ &	$            0.27$ &	$         -6.7385$ &	$         2972.76$\\
	$            5.55$ &	$           31.65$ &	$           22.60$ &	$           29569$ &	$             0.2$ &	$            0.29$ &	$         -7.3288$ &	$         3437.10$\\
	$            5.55$ &	$           31.65$ &	$           22.60$ &	$           29569$ &	$             0.5$ &	$            0.29$ &	$         -6.9958$ &	$         3473.92$\\
	$            5.55$ &	$           31.65$ &	$           22.60$ &	$           29569$ &	$             1.0$ &	$            0.29$ &	$         -6.5934$ &	$         2439.75$\\
	$            5.13$ &	$           21.11$ &	$           13.37$ &	$           30231$ &	$             0.2$ &	$            0.17$ &	$         -8.0491$ &	$         3387.11$\\
	$            5.13$ &	$           21.11$ &	$           13.37$ &	$           30231$ &	$             0.5$ &	$            0.17$ &	$         -7.6460$ &	$         3603.45$\\
	$            5.13$ &	$           21.11$ &	$           13.37$ &	$           30231$ &	$             1.0$ &	$            0.17$ &	$         -7.3292$ &	$         2865.80$\\
	$            4.63$ &	$           16.57$ &	$            7.39$ &	$           30488$ &	$             0.2$ &	$            0.07$ &	$         -9.2680$ &	$         2995.31$\\
	$            4.63$ &	$           16.57$ &	$            7.39$ &	$           30488$ &	$             0.5$ &	$            0.07$ &	$         -8.5506$ &	$         3609.03$\\
	$            4.63$ &	$           16.57$ &	$            7.39$ &	$           30488$ &	$             1.0$ &	$            0.07$ &	$         -8.2760$ &	$         3976.81$\\
	$            5.59$ &	$           34.27$ &	$           22.20$ &	$           30504$ &	$             0.2$ &	$            0.30$ &	$         -7.4192$ &	$         3687.45$\\
	$            5.59$ &	$           34.27$ &	$           22.20$ &	$           30504$ &	$             0.5$ &	$            0.30$ &	$         -6.6778$ &	$         2092.16$\\
	$            5.59$ &	$           34.27$ &	$           22.20$ &	$           30504$ &	$             1.0$ &	$            0.30$ &	$         -6.6745$ &	$         2608.41$\\
	$            5.18$ &	$           23.17$ &	$           13.69$ &	$           30737$ &	$             0.2$ &	$            0.17$ &	$         -8.0130$ &	$         3566.44$\\
	$            5.18$ &	$           23.17$ &	$           13.69$ &	$           30737$ &	$             0.5$ &	$            0.17$ &	$         -7.4590$ &	$         3091.28$\\
	$            5.18$ &	$           23.17$ &	$           13.69$ &	$           30737$ &	$             1.0$ &	$            0.17$ &	$         -7.3701$ &	$         3716.56$\\
	$            5.61$ &	$           37.00$ &	$           22.03$ &	$           31009$ &	$             0.2$ &	$            0.29$ &	$         -7.4455$ &	$         3762.63$\\
	$            5.61$ &	$           37.00$ &	$           22.03$ &	$           31009$ &	$             0.5$ &	$            0.29$ &	$         -6.8865$ &	$         2936.07$\\
	$            5.61$ &	$           37.00$ &	$           22.03$ &	$           31009$ &	$             1.0$ &	$            0.29$ &	$         -6.5546$ &	$         2349.45$\\
	$            4.73$ &	$           18.14$ &	$            7.73$ &	$           31524$ &	$             0.2$ &	$            0.08$ &	$         -9.2122$ &	$         2925.22$\\
	$            4.73$ &	$           18.14$ &	$            7.73$ &	$           31524$ &	$             0.5$ &	$            0.08$ &	$         -8.4926$ &	$         4140.30$\\
	$            4.73$ &	$           18.14$ &	$            7.73$ &	$           31524$ &	$             1.0$ &	$            0.08$ &	$         -8.1367$ &	$         3740.38$\\
	$            5.24$ &	$           24.94$ &	$           13.88$ &	$           31689$ &	$             0.2$ &	$            0.19$ &	$         -7.9754$ &	$         3951.62$\\
	$            5.24$ &	$           24.94$ &	$           13.88$ &	$           31689$ &	$             0.5$ &	$            0.19$ &	$         -7.4511$ &	$         3263.31$\\
	$            5.24$ &	$           24.94$ &	$           13.88$ &	$           31689$ &	$             1.0$ &	$            0.19$ &	$         -7.2669$ &	$         3392.29$\\
	$            5.64$ &	$           39.33$ &	$           21.69$ &	$           31913$ &	$             0.2$ &	$            0.30$ &	$         -7.3891$ &	$         3921.73$\\
	$            5.64$ &	$           39.33$ &	$           21.69$ &	$           31913$ &	$             0.5$ &	$            0.30$ &	$         -6.8835$ &	$         3144.53$\\
	$            5.64$ &	$           39.33$ &	$           21.69$ &	$           31913$ &	$             1.0$ &	$            0.30$ &	$         -6.6351$ &	$         2735.57$\\
	$            4.82$ &	$           19.96$ &	$            8.11$ &	$           32522$ &	$             0.2$ &	$            0.09$ &	$         -8.9941$ &	$         2892.27$\\
	$            4.82$ &	$           19.96$ &	$            8.11$ &	$           32522$ &	$             0.5$ &	$            0.09$ &	$         -8.4004$ &	$         4461.07$\\
	$            4.82$ &	$           19.96$ &	$            8.11$ &	$           32522$ &	$             1.0$ &	$            0.09$ &	$         -8.0413$ &	$         3784.51$\\
	$            5.31$ &	$           26.99$ &	$           14.11$ &	$           32573$ &	$             0.2$ &	$            0.20$ &	$         -7.9585$ &	$         3564.44$\\
	$            5.31$ &	$           26.99$ &	$           14.11$ &	$           32573$ &	$             0.5$ &	$            0.20$ &	$         -7.1971$ &	$         2443.79$\\
	$            5.31$ &	$           26.99$ &	$           14.11$ &	$           32573$ &	$             1.0$ &	$            0.20$ &	$         -7.1019$ &	$         2720.15$\\
	$            5.70$ &	$           40.96$ &	$           21.14$ &	$           33326$ &	$             0.2$ &	$            0.32$ &	$         -7.1702$ &	$         4078.63$\\
	$            5.70$ &	$           40.96$ &	$           21.14$ &	$           33326$ &	$             0.5$ &	$            0.32$ &	$         -6.6501$ &	$         2592.01$\\
	$            5.70$ &	$           40.96$ &	$           21.14$ &	$           33326$ &	$             1.0$ &	$            0.32$ &	$         -6.3665$ &	$         2400.43$\\
	$            4.91$ &	$           22.03$ &	$            8.52$ &	$           33383$ &	$             0.2$ &	$            0.10$ &	$         -8.7594$ &	$         3003.33$\\
	$            4.91$ &	$           22.03$ &	$            8.52$ &	$           33383$ &	$             0.5$ &	$            0.10$ &	$         -8.1931$ &	$         4596.93$\\
	$            4.91$ &	$           22.03$ &	$            8.52$ &	$           33383$ &	$             1.0$ &	$            0.10$ &	$         -8.1205$ &	$         5411.75$\\
	$            5.37$ &	$           29.19$ &	$           14.34$ &	$           33487$ &	$             0.2$ &	$            0.21$ &	$         -7.7219$ &	$         3667.92$\\
	$            5.37$ &	$           29.19$ &	$           14.34$ &	$           33487$ &	$             0.5$ &	$            0.21$ &	$         -7.4443$ &	$         4501.69$\\
	$            5.37$ &	$           29.19$ &	$           14.34$ &	$           33487$ &	$             1.0$ &	$            0.21$ &	$         -7.1115$ &	$         3415.22$\\
	$            5.01$ &	$           24.26$ &	$            8.94$ &	$           34419$ &	$             0.2$ &	$            0.11$ &	$         -8.3905$ &	$         3215.37$\\
	$            5.01$ &	$           24.26$ &	$            8.94$ &	$           34419$ &	$             0.5$ &	$            0.11$ &	$         -8.0308$ &	$         4424.16$\\
	$            5.01$ &	$           24.26$ &	$            8.94$ &	$           34419$ &	$             1.0$ &	$            0.11$ &	$         -7.8713$ &	$         5481.06$\\
	$            5.44$ &	$           31.30$ &	$           14.51$ &	$           34638$ &	$             0.2$ &	$            0.23$ &	$         -7.4983$ &	$         3891.42$\\
	$            5.44$ &	$           31.30$ &	$           14.51$ &	$           34638$ &	$             0.5$ &	$            0.23$ &	$         -7.2115$ &	$         4251.11$\\
	$            5.44$ &	$           31.30$ &	$           14.51$ &	$           34638$ &	$             1.0$ &	$            0.23$ &	$         -6.8466$ &	$         3008.50$\\
	$            5.75$ &	$           42.98$ &	$           20.68$ &	$           34654$ &	$             0.2$ &	$            0.34$ &	$         -6.8965$ &	$         3794.50$\\
	$            5.75$ &	$           42.98$ &	$           20.68$ &	$           34654$ &	$             0.5$ &	$            0.34$ &	$         -6.5019$ &	$         2731.52$\\
	$            5.75$ &	$           42.98$ &	$           20.68$ &	$           34654$ &	$             1.0$ &	$            0.34$ &	$         -6.1938$ &	$         2377.51$\\
	$            5.10$ &	$           26.65$ &	$            9.37$ &	$           35531$ &	$             0.2$ &	$            0.12$ &	$         -8.1450$ &	$         3641.13$\\
	$            5.10$ &	$           26.65$ &	$            9.37$ &	$           35531$ &	$             0.5$ &	$            0.12$ &	$         -7.8871$ &	$         5044.18$\\
	$            5.10$ &	$           26.65$ &	$            9.37$ &	$           35531$ &	$             1.0$ &	$            0.12$ &	$         -7.5994$ &	$         5386.62$\\
	$            5.50$ &	$           33.82$ &	$           14.74$ &	$           35644$ &	$             0.2$ &	$            0.25$ &	$         -7.2806$ &	$         3988.51$\\
	$            5.50$ &	$           33.82$ &	$           14.74$ &	$           35644$ &	$             0.5$ &	$            0.25$ &	$         -6.9815$ &	$         3959.17$\\
	$            5.50$ &	$           33.82$ &	$           14.74$ &	$           35644$ &	$             1.0$ &	$            0.25$ &	$         -6.6525$ &	$         2811.78$\\
	$            5.78$ &	$           45.54$ &	$           20.33$ &	$           35747$ &	$             0.2$ &	$            0.35$ &	$         -6.7298$ &	$         3745.48$\\
	$            5.78$ &	$           45.54$ &	$           20.33$ &	$           35747$ &	$             0.5$ &	$            0.35$ &	$         -6.3424$ &	$         2517.89$\\
	$            5.78$ &	$           45.54$ &	$           20.33$ &	$           35747$ &	$             1.0$ &	$            0.35$ &	$         -5.9907$ &	$         2136.93$\\
	$            5.56$ &	$           36.53$ &	$           14.97$ &	$           36673$ &	$             0.2$ &	$            0.26$ &	$         -7.1085$ &	$         4047.17$\\
	$            5.56$ &	$           36.53$ &	$           14.97$ &	$           36673$ &	$             0.5$ &	$            0.26$ &	$         -6.7414$ &	$         3307.20$\\
	$            5.56$ &	$           36.53$ &	$           14.97$ &	$           36673$ &	$             1.0$ &	$            0.26$ &	$         -6.4660$ &	$         2671.89$\\
	$            5.20$ &	$           29.09$ &	$            9.79$ &	$           36826$ &	$             0.2$ &	$            0.14$ &	$         -7.7563$ &	$         3857.78$\\
	$            5.20$ &	$           29.09$ &	$            9.79$ &	$           36826$ &	$             0.5$ &	$            0.14$ &	$         -7.5491$ &	$         5104.77$\\
	$            5.20$ &	$           29.09$ &	$            9.79$ &	$           36826$ &	$             1.0$ &	$            0.14$ &	$         -7.2386$ &	$         4389.04$\\
	$            5.83$ &	$           47.94$ &	$           19.92$ &	$           37070$ &	$             0.2$ &	$            0.37$ &	$         -6.5079$ &	$         3189.74$\\
	$            5.83$ &	$           47.94$ &	$           19.92$ &	$           37070$ &	$             0.5$ &	$            0.37$ &	$         -6.1263$ &	$         2201.22$\\
	$            5.83$ &	$           47.94$ &	$           19.92$ &	$           37070$ &	$             1.0$ &	$            0.37$ &	$         -5.8465$ &	$         2131.97$\\
	$            5.63$ &	$           39.07$ &	$           15.13$ &	$           38003$ &	$             0.2$ &	$            0.29$ &	$         -6.8484$ &	$         3733.72$\\
	$            5.63$ &	$           39.07$ &	$           15.13$ &	$           38003$ &	$             0.5$ &	$            0.29$ &	$         -6.5024$ &	$         2880.15$\\
	$            5.63$ &	$           39.07$ &	$           15.13$ &	$           38003$ &	$             1.0$ &	$            0.29$ &	$         -6.2001$ &	$         2368.80$\\
	$            5.30$ &	$           31.76$ &	$           10.23$ &	$           38151$ &	$             0.2$ &	$            0.17$ &	$         -7.5341$ &	$         4126.77$\\
	$            5.30$ &	$           31.76$ &	$           10.23$ &	$           38151$ &	$             0.5$ &	$            0.17$ &	$         -7.2155$ &	$         4545.51$\\
	$            5.30$ &	$           31.76$ &	$           10.23$ &	$           38151$ &	$             1.0$ &	$            0.17$ &	$         -6.9668$ &	$         3881.23$\\
	$            5.88$ &	$           51.44$ &	$           19.48$ &	$           38520$ &	$             0.2$ &	$            0.39$ &	$         -6.3332$ &	$         2964.94$\\
	$            5.88$ &	$           51.44$ &	$           19.48$ &	$           38520$ &	$             0.5$ &	$            0.39$ &	$         -5.9776$ &	$         2123.92$\\
	$            5.88$ &	$           51.44$ &	$           19.48$ &	$           38520$ &	$             1.0$ &	$            0.39$ &	$         -5.7181$ &	$         2109.59$\\
	$            5.71$ &	$           41.62$ &	$           15.26$ &	$           39507$ &	$             0.2$ &	$            0.32$ &	$         -6.5947$ &	$         3369.99$\\
	$            5.71$ &	$           41.62$ &	$           15.26$ &	$           39507$ &	$             0.5$ &	$            0.32$ &	$         -6.2538$ &	$         2459.56$\\
	$            5.71$ &	$           41.62$ &	$           15.26$ &	$           39507$ &	$             1.0$ &	$            0.32$ &	$         -5.9873$ &	$         2248.02$\\
	$            5.42$ &	$           34.17$ &	$           10.61$ &	$           40062$ &	$             0.2$ &	$            0.20$ &	$         -7.2082$ &	$         4105.09$\\
	$            5.42$ &	$           34.17$ &	$           10.61$ &	$           40062$ &	$             0.5$ &	$            0.20$ &	$         -6.8606$ &	$         3905.13$\\
	$            5.42$ &	$           34.17$ &	$           10.61$ &	$           40062$ &	$             1.0$ &	$            0.20$ &	$         -6.5964$ &	$         2981.86$\\
	$            5.95$ &	$           58.28$ &	$           18.91$ &	$           40702$ &	$             0.2$ &	$            0.40$ &	$         -6.1306$ &	$         2838.93$\\
	$            5.95$ &	$           58.28$ &	$           18.91$ &	$           40702$ &	$             0.5$ &	$            0.40$ &	$         -5.8252$ &	$         2297.17$\\
	$            5.95$ &	$           58.28$ &	$           18.91$ &	$           40702$ &	$             1.0$ &	$            0.40$ &	$         -5.6096$ &	$         2302.44$\\
	$            5.83$ &	$           49.10$ &	$           15.83$ &	$           41486$ &	$             0.2$ &	$            0.36$ &	$         -6.3146$ &	$         3106.91$\\
	$            5.83$ &	$           49.10$ &	$           15.83$ &	$           41486$ &	$             0.5$ &	$            0.36$ &	$         -6.0056$ &	$         2500.96$\\
	$            5.83$ &	$           49.10$ &	$           15.83$ &	$           41486$ &	$             1.0$ &	$            0.36$ &	$         -5.7866$ &	$         2384.98$\\
	$            5.52$ &	$           37.26$ &	$           11.08$ &	$           41540$ &	$             0.2$ &	$            0.23$ &	$         -6.9302$ &	$         3845.47$\\
	$            5.52$ &	$           37.26$ &	$           11.08$ &	$           41540$ &	$             0.5$ &	$            0.23$ &	$         -6.5930$ &	$         3445.92$\\
	$            5.52$ &	$           37.26$ &	$           11.08$ &	$           41540$ &	$             1.0$ &	$            0.23$ &	$         -6.3529$ &	$         2795.81$\\
	$            6.00$ &	$           66.85$ &	$           18.47$ &	$           42551$ &	$             0.2$ &	$            0.40$ &	$         -6.0439$ &	$         3197.22$\\
	$            6.00$ &	$           66.85$ &	$           18.47$ &	$           42551$ &	$             0.5$ &	$            0.40$ &	$         -5.7596$ &	$         2715.89$\\
	$            6.00$ &	$           66.85$ &	$           18.47$ &	$           42551$ &	$             1.0$ &	$            0.40$ &	$         -5.5630$ &	$         2543.34$\\
	$            5.93$ &	$           58.99$ &	$           16.57$ &	$           42942$ &	$             0.2$ &	$            0.38$ &	$         -6.1376$ &	$         3152.19$\\
	$            5.93$ &	$           58.99$ &	$           16.57$ &	$           42942$ &	$             0.5$ &	$            0.38$ &	$         -5.8577$ &	$         2782.45$\\
	$            5.93$ &	$           58.99$ &	$           16.57$ &	$           42942$ &	$             1.0$ &	$            0.38$ &	$         -5.6784$ &	$         2647.87$\\
	$            5.69$ &	$           45.99$ &	$           12.31$ &	$           43419$ &	$             0.2$ &	$            0.28$ &	$         -6.5800$ &	$         3658.03$\\
	$            5.69$ &	$           45.99$ &	$           12.31$ &	$           43419$ &	$             0.5$ &	$            0.28$ &	$         -6.2717$ &	$         3399.20$\\
	$            5.69$ &	$           45.99$ &	$           12.31$ &	$           43419$ &	$             1.0$ &	$            0.28$ &	$         -6.0704$ &	$         3057.32$\\
	$            5.84$ &	$           58.13$ &	$           13.84$ &	$           44616$ &	$             0.2$ &	$            0.31$ &	$         -6.3267$ &	$         3667.48$\\
	$            5.84$ &	$           58.13$ &	$           13.84$ &	$           44616$ &	$             0.5$ &	$            0.31$ &	$         -6.0451$ &	$         3517.46$\\
	$            5.84$ &	$           58.13$ &	$           13.84$ &	$           44616$ &	$             1.0$ &	$            0.31$ &	$         -5.8573$ &	$         3152.44$\\
    \end{longtable}


\end{document}